\documentclass[%
 aip,
 amsmath,amssymb,
 reprint,%
]{revtex4-1}

\usepackage{xcolor}
\usepackage{graphicx}
\usepackage{dcolumn}
\usepackage{bm}
\usepackage{siunitx}
\DeclareSIUnit{\atm}{atm}
\usepackage{subcaption}

\usepackage[utf8]{inputenc}
\usepackage[T1]{fontenc}
\usepackage{mathptmx}
\usepackage{etoolbox}

\makeatletter
\def\@email#1#2{%
 \endgroup
 \patchcmd{\titleblock@produce}
  {\frontmatter@RRAPformat}
  {\frontmatter@RRAPformat{\produce@RRAP{*#1\href{mailto:#2}{#2}}}\frontmatter@RRAPformat}
  {}{}
}%
\makeatother
\begin{document}

\preprint{AIP/123-QED}

\title{Simulated Solvation Dynamics of a Photoexcited Organic Molecule and the Associated Time-Resolved X-ray Scattering Signal}
\author{S. Petry}
\affiliation{Department of Chemical and Pharmaceutical Sciences, University of Trieste, 34127 Trieste, Italy}
\email{simon.petry@units.it}
\author{M. Kristinsson}%
\affiliation{Department of Chemistry, Technical University of Denmark, Kongens Lyngby, Denmark.}
\author{A. O. Dohn}
\affiliation{Department of Physics, Technical University of Denmark, Kongens Lyngby, Denmark.}
\author{K. B. M{\o}ller}
\affiliation{Department of Chemistry, Technical University of Denmark, Kongens Lyngby, Denmark.}
\author{E. Selenius}
\affiliation{Science Institute and Faculty of Physical Sciences, University of Iceland, 107 Reykjavík, Iceland}
\author{G. Levi}
\affiliation{Department of Chemical and Pharmaceutical Sciences, University of Trieste, 34127 Trieste, Italy}
\affiliation{Science Institute and Faculty of Physical Sciences, University of Iceland, 107 Reykjavík, Iceland}
\email{gianluca.levi@units.it}

\date{\today}

\begin{abstract}
Solvation shell reorganization is a key component of the dynamics after photoexcitation of a solvated molecule, especially for charge-transfer excitation in polar solvents.
These structural dynamics can be directly observed with time-resolved X-ray solution scattering (TR-XSS) experiments. However, disentangling the scattering signal of solvation-shell reorganization can be challenging, especially for systems containing only light atoms.
Here, nonequilibrium molecular dynamics simulations of a hemithioindigo chromophore in acetonitrile are used to investigate the solvation dynamics upon photoinduced charge transfer as well as vibrational excitation and cooling, and to characterize the corresponding time-resolved X-ray scattering signals.
It is found that charge transfer induces rapid reorientation of the first solvation shell (\(0.15\)--\(0.2~\mathrm{ps}\)) followed by slower reorientational and solvent redistribution within and between shells (\(1\)--\(2~\mathrm{ps}\)).
An analysis of the volume around the solute excluded to the solvent reveals an initial contraction of the solvation shell, followed by transient rebound within the first \(0.4~\mathrm{ps}\), and then slower contraction.
Instantaneous vibrational excitation expands the excluded volume, reaching a maximum near \(0.4~\mathrm{ps}\), followed by contraction during cooling with biexponential relaxation. Low-$q$ features in both the calculated solute–solvent and solvent–solvent scattering signals reflect rearrangements within the first solvation shell and redistribution across shells, while higher-$q$ solute–solvent features more closely follow local first-shell restructuring. These results show that TR-XSS can provide information on both local solvent rearrangements and redistribution across solvation shells and demonstrate that interpreting the scattering signals requires accounting for changes in solvent packing induced by vibrational heating and cooling of the solute.
\end{abstract}

\maketitle
 
\section{\label{sec:introduction}Introduction}

Photoinduced charge transfer redistributes the electronic charge density of a chromophore, changing its electrostatic interaction with the surrounding solvent.
The solvent responds through both electronic polarization and nuclear motion.
In polar solvents, a prominent component of the solvent response is the dipolar reorganization of the solvent molecules.\cite{Bagchi_2010,Fleming_1989}
By modifying the relative stability of electronic states and the free-energy barriers, polar solvation can influence charge localization, excited-state relaxation pathways, and ultimately photochemical outcomes.\cite{Dereka_2020,Kumpulainen_2017_review,Rondi_2015}
Such solvation processes have been extensively investigated using time-resolved optical spectroscopy, particularly measurements of the dynamic Stokes shift, which mainly report on the  relaxation of the solvation free energy.\cite{Kumpulainen_2017_review,Horng_1995}
These studies have identified two different regimes: a fast response occurring within tens to a few hundred femtoseconds, commonly termed ``inertial''\footnote{Sometimes, the term ``inertial'' has been used in the literature in the broader sense of short-time dynamics. The solvent motions need not be inertial in the strictly mechanical sense and may be influenced by the electrostatic torque exerted by the change in the solute electric field.\cite{Rey_2020}}, and slower relaxation spanning fractions of a picosecond to several hundred picoseconds depending on the solvent.\cite{Horng_1995,Jimenez_1994,Rosenthal_1991}
Optical measurements provide sensitive information about energetic relaxation and its time scales. However, they provide only indirect information about the underlying changes in the position and orientation of solvent molecules. Molecular dynamics (MD) simulations have provided a microscopic interpretation of the two regimes, associating the fast response mainly with reorientation and librational motion of solvent molecules, and the slower response with large amplitude rearrangements of the solvation environment, commonly referred to as ``diffusive''.\cite{Rey_2015,Ingrosso_2005,Kumar_1995,Maroncelli_1991}

Time-resolved X-ray solution scattering (TR-XSS)\cite{Lee_2024,Choi_2022} offers a complementary approach, as the scattering signal is directly sensitive to changes in all atom-pair distances.
Previous studies have resolved solvation-shell rearrangements around photoexcited molecules containing  halogen\cite{Markmann_2024,Nimmrich_2023,Vester_2022,Heo_2022,Lee_2013} and metal atoms,\cite{vanDriel_2016,Haldrup_2016} including coherent solvent motions coupled to intramolecular charge transfer,\cite{Biasin_2021} and more recently, the solvent reorganization in proton-coupled electron transfer.\cite{Kahraman_2026}
These studies benefited from the strong scattering of heavy solute atoms, which enhances sensitivity to changes in solute--solvent distances.
Recently, the first direct atomic-scale observation by TR-XSS of polar solvation dynamics around a photoexcited organic chromophore containing only light atoms was reported for an hemithioindigo compound (HTI) in acetonitrile.\cite{Mitterer_2026}
The measurements identified a transient scattering signature of the solvation shell that forms after photoinduced charge transfer with a time constant of approximately \(0.3~\mathrm{ps}\) and decays over approximately \(3~\mathrm{ps}\) as the chromophore returns to its ground state. Comparison with MD simulations attributed the solvent reorganization predominantly to the response to photoinduced charge transfer rather than structural changes of the solute.
More recently, the first TR-XSS observation of solute structural dynamics in a small organic molecule containing only light atoms was reported for azobenzene photoisomerization in methanol.\cite{Kim_2026}
These recent studies represent important advances in the direct observation of photoinduced dynamics of systems including only light atoms, but also highlight the remaining challenges of associating measured scattering signals of solvated organic molecules with specific molecular motions.

Photoexcitation triggers several coupled processes that contribute simultaneously to the measured scattering signal.
In addition to solute structural changes and the response of the surrounding solvation shell to photoexcitation, the solute is vibrationally excited as a result of the initial excitation as well as internal conversion processes. This can modify the local solvent packing, producing an additional scattering response.\cite{Mitterer_2026} Eventually, relaxation of the initially nonequilibrium solvation shell and vibrational energy relaxation of the solute release energy into the solvent\cite{Rey_2015}, leading to solvent heating. The bulk scattering response associated with the solvent heating typically dominates the signal of weakly scattering organic solutes and poses a considerable problem in the interpretation of the signal.\cite{Kim_2026,Mitterer_2026}
Other contributions, such as solvent motions associated with the optical Kerr effect, can further influence the early-time signal.\cite{Ki_2021}
All these contributions can overlap in momentum-transfer ($q$) space and evolve on similar time scales, making it difficult to disentangle the signatures of solute and solvation dynamics. In the case of the HTI study, the experiments could not separately resolve the orientational and translational rearrangements of acetonitrile molecules, or determine the time scales of these processes.\cite{Mitterer_2026}

MD simulations can connect structural changes to X-ray scattering signals through atom-pair radial distribution functions (RDFs).\cite{Dohn_2023,Dohn_2015}
When used for the analysis of TR-XSS data, these simulations often employ classical force fields in which electrostatic interactions are represented by fixed-value atomic partial charges.\cite{Kahraman_2026, Mara_2025, Lenzen_2025, Hansen_2025, Nimmrich_2023, Oang_2023, Vester_2022, Biasin_2021, Ki_2021, Panman_2020, Biasin_2016, Haldrup_2016, Lee_2013}
Their comparatively low computational cost enables the extensive sampling needed to converge weak difference signals and compare models with different solute geometries and charge distributions.
Usually, equilibrium simulations of individual electronic states are used to model solvation structures and the corresponding scattering signatures, although they do not directly reveal the pathway or time scales of solvent reorganization.\cite{Kim_2026,Mitterer_2026}
The transient response can instead be followed through ensembles of nonequilibrium trajectories, commonly initiated by instantaneously changing the solute partial charges to represent electronic excitation.\cite{Papai_2019,Rey_2015,Ingrosso_2005}
Such simulations provide useful information on polar solvation dynamics when nuclear reorganization dominates the solvent response.
They have also supported the interpretation of TR-XSS signals.\cite{Biasin_2021}
However, these studies focused on the solvent response to electronic excitation, neglecting solute vibrational excitation and cooling.

In the present work, we use nonequilibrium MD simulations of HTI in acetonitrile to investigate two key processes following photoexcitation: solvation dynamics driven by photoinduced charge transfer, and vibrational heating and subsequent cooling of the solute.
By examining these processes separately, we distinguish their effect on the solvation structure and their contributions to the TR-XSS signal.

The trajectories are analyzed using time-resolved spatial distribution functions (SDFs), which characterize the solvent structure in terms of angular orientation and radial distance with respect to the solute (see Sec.~\ref{SEC:SDFs}). This allows solvent
reorientation and translational restructuring to be followed separately.
We also track changes in the excluded volume, defined as the volume surrounding the solute that is inaccessible to solvent molecules, relating these changes to the scattering signal, and thereby extending previous investigations of the excluded volume effect in static solution scattering to the nonequilibrium regime.\cite{Dohn_2023, Thompson_2024}

The charge-transfer simulations reveal rapid solvent reorientation, predominantly within the first solvation shell, on a time scale of approximately \(0.15\)--\(0.2~\mathrm{ps}\), accompanied by an initial contraction of the excluded volume.
Following a brief rebound, a slower solvent rearrangement produces further contraction on the picosecond time scale.
Vibrational excitation causes a rapid expansion of the excluded volume, followed by contraction as the solute cools and transfers energy to the solvent on time scales of several to tens of picoseconds.
Connecting these structural changes to the calculated scattering signals provides a basis for interpreting the signal in different \(q\) ranges and distinguishing polar solvation dynamics from the response to vibrational excitation in TR-XSS measurements of organic molecules in solution.

\section{\label{sec:compdetails}Methodology and computational settings}
\subsection{Molecular model}
The simulated system consists of one hemithioindigo compound where the stilbene moiety is a julolidine group (hereafter referred to as HTI, Fig.~\ref{fig:SDF_definition}) solvated by 717 acetonitrile molecules in a cubic periodic simulation cell.

Equilibrium bond lengths and angles for the force field of HTI are taken from the ground-state geometry optimized using density functional theory (DFT) with the CAM-B3LYP functional\cite{Yanai_2004} and the ma-def2-TZVP basis set.\cite{Zheng_2011,Weigend_2005}
Atomic partial charges are obtained by CHELPG fitting to the electrostatic potential,\cite{Breneman_1990} with acetonitrile represented by the conductor-like polarizable continuum model (CPCM).\cite{Cossi_2003,Barone_1998}
The remaining bonded and Lennard--Jones parameters are based on the generalized Amber force field (GAFF).\cite{Gaff_2004}
Torsional force constants for the C--C single bond connecting the stilbene and julolidine groups are increased to restrain the large-amplitude twisting, as preliminary calculations with the original force field showed large deviations from the DFT torsional energy profile.
The charge-transfer excited state is represented by a second HTI parameter set, where the partial charges are obtained from an excited-state CAM-B3LYP/ma-def2-TZVP calculation using an orbital-optimized, variational density functional approach.\cite{Restaino_2026,Schmerwitz_2026,Levi_2020}
All bonded and Lennard--Jones parameters are retained from the ground-state model. Consequently, the ground- and excited-state force fields of HTI differ only through the partial charges.

The DFT calculations were performed using version 6.0 of the ORCA software \cite{Neese2012, Neese2022}.

\subsection{Equilibrium molecular dynamics}
The system is initially equilibrated in the ground state in the NPT ensemble for \(5~\si{ns}\) at a temperature of \(298.15~\si{K}\) and a pressure of \(1~\si{atm}\). The resulting equilibrium cubic simulation box has a side length of approximately \(40.1~\text{\AA}\). Subsequent equilibrium sampling is performed in the NVT ensemble at \(298.15~\mathrm{K}\) using a Langevin middle integrator (friction coefficient of \(1.0~\mathrm{ps}^{-1}\)) with a time step of \(1~\mathrm{fs}\). Electrostatic interactions are treated using particle-mesh Ewald electrostatics with a real-space cutoff of \(1.0~\mathrm{nm}\). No bond constraints are applied, and the center-of-mass motion of the system is removed during the dynamics. The system is propagated in the NVT ensemble for a total of \(405~\mathrm{ns}\), comprising \(5~\mathrm{ns}\) equilibration followed by \(400~\mathrm{ns}\) equilibrium sampling. Coordinates, energy, and temperature are recorded every \(1~\mathrm{ps}\).
All molecular dynamics simulations are performed with OpenMM 8.2.0\cite{Openmm_2024}.

\subsection{Nonequilibrium molecular dynamics}

\subsubsection{Photoinduced charge transfer}

A total of 9,000 configurations are extracted from the equilibrated NVT trajectory, spaced by at least \(40~\mathrm{ps}\).
For each configuration, velocities are newly assigned from a Maxwell--Boltzmann distribution at \(298.15~\mathrm{K}\).
Each configuration is then propagated for \(20~\mathrm{ps}\) in the ground state in the NVE ensemble using a leap-frog Verlet integrator with a time step of \(0.5~\mathrm{fs}\).
The configurations obtained at the end of this ground-state propagation are taken as the initial frames for the nonequilibrium simulations.

The photoexcitation is represented by an instantaneous replacement of the ground-state HTI parameter set with the charge-transfer-state parameter set.
Atomic positions, velocities, simulation-box vectors, and all remaining force-field parameters remain unchanged during this switch. From each initial frame, one nonequilibrium trajectory is generated by propagating the system for \(5~\mathrm{ps}\) in the NVE ensemble using the charge-transfer state force field and a time step of \(0.5~\mathrm{fs}\).
All 9,000 nonequilibrium trajectories are included in the final analysis, unless otherwise stated.
The trajectories are saved with the following intervals:

\begin{equation}
\Delta t_{\mathrm{save}} =
\begin{cases}
1~\mathrm{fs}, & 0 < t \leq 0.5~\mathrm{ps},\\
5~\mathrm{fs}, & 0.5 < t \leq 2.5~\mathrm{ps},\\
10~\mathrm{fs}, & 2.5 < t \leq 5.0~\mathrm{ps}.
\end{cases}
\end{equation}

\subsubsection{Vibrational excitation}
A second set of 5.000 nonequilibrium simulations is generated from configurations extracted from a separate equilibrated NVT trajectory using the ground-state molecular model.
Following the same procedure as for the photoinduced charge-transfer simulations, each configuration is propagated for \(20~\mathrm{ps}\) in the ground state in the NVE ensemble before vibrational excitation. 
For the excitation, the velocities of the solute atoms are resampled from a Maxwell-Boltzmann distribution and instantaneously changed to increase the solute kinetic energy by 2.64 eV.
The resulting trajectories are propagated for \(60~\mathrm{ps}\) in the NVE ensemble with a time step of \(0.5~\mathrm{fs}\). The nonequilibrium trajectories are saved with the following intervals:

\begin{equation}
\Delta t_{\mathrm{save}} =
\begin{cases}
1~\mathrm{fs}, & 0 < t \leq 0.2~\mathrm{ps},\\
5~\mathrm{fs}, & 0.2 < t \leq 8~\mathrm{ps},\\
1~\mathrm{ps}, & 8 < t \leq 60~\mathrm{ps}.
\end{cases}
\end{equation}

\subsection{Structural analysis and scattering calculations}

\subsubsection{Spatial distribution functions}
\label{SEC:SDFs}

\begin{figure}[t]
    \centering
    \includegraphics[width=0.8\columnwidth]{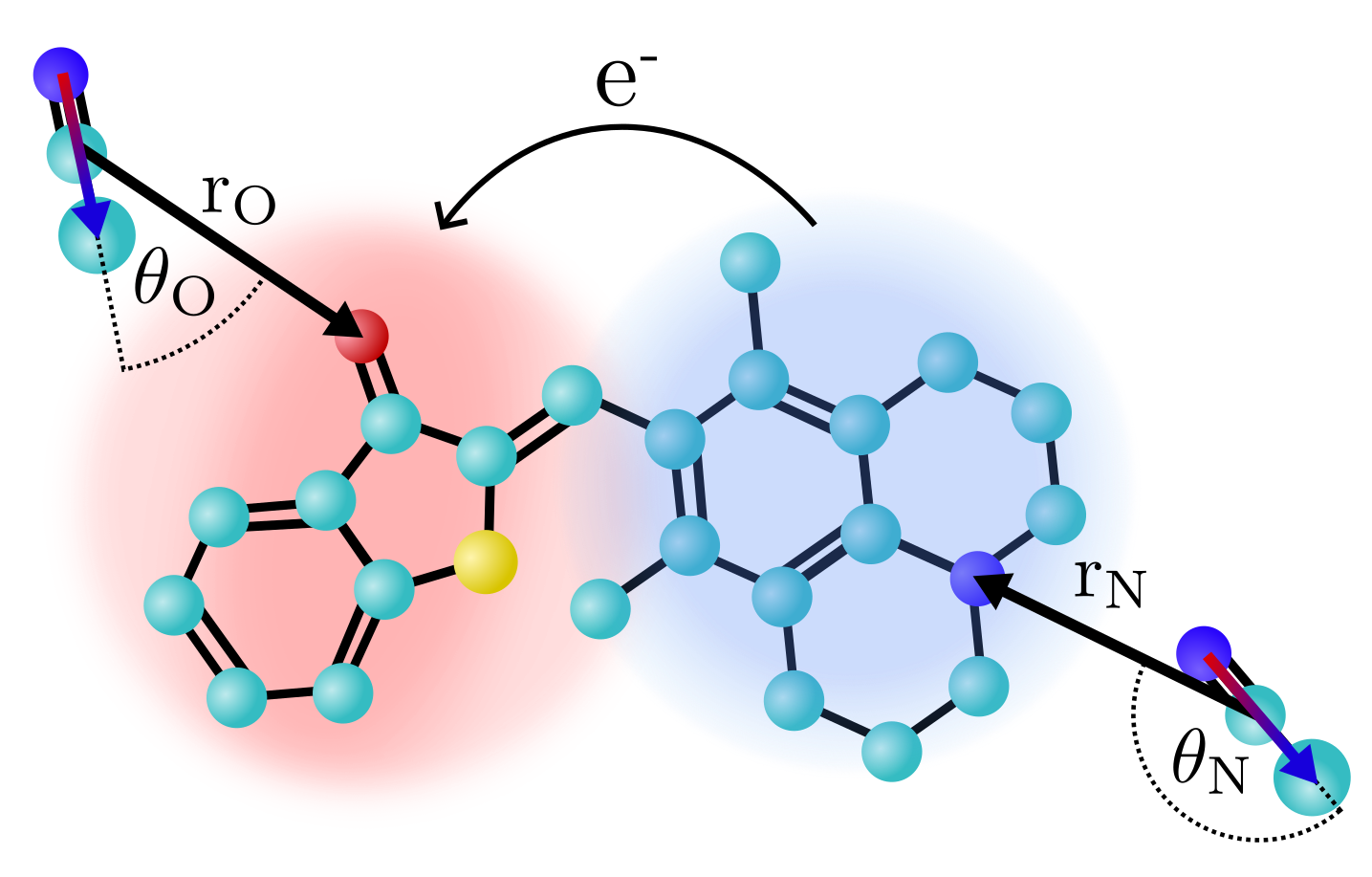}
    \caption{
    The structure of the hemithioindigo compound \mbox{HTI} indicating the direction of the photoinduced charge transfer and a geometric definition of the radial and angular coordinates used for the spatial distribution functions. Hydrogen atoms are omitted for clarity.
    For a selected \mbox{HTI} reference atom \(X\), the radial coordinate \(r_X\) is the distance from the midpoint between the nitrogen atom ($N$) and the methyl carbon ($C_{\mathrm{Me}}$) of acetonitrile molecule \(j\) to \(X\), while \(\theta_X\) is the angle between the acetonitrile \(N\!\rightarrow C_{\mathrm{Me}}\) orientation vector and the vector from the \(N\!-C_{\mathrm{Me}}\) midpoint toward \(X\).
    The definitions are shown for \(X=N_{\mathrm{HTI}}\) and \(X=O_{\mathrm{HTI}}\).}
    \label{fig:SDF_definition}
\end{figure}

The local solvent structure around selected HTI atoms \(X\) is described by spatial distribution functions (SDFs), \(s_X(r,t,\theta)\), which characterize the joint distribution of distances and molecular orientations at delay time \(t\).
The distance \(r\) is measured from \(X\) to the midpoint between the nitrogen atom \(N\) and methyl carbon \(C_{\mathrm{Me}}\) of an acetonitrile molecule.
The angle \(\theta\) is defined between the acetonitrile \(N\!\rightarrow C_{\mathrm{Me}}\) axis and the vector pointing from this midpoint toward \(X\), as illustrated in Fig.~\ref{fig:SDF_definition}.
According to this convention, \(\theta=0^\circ\) denotes alignment with the methyl end pointing toward \(X\), and \(\theta=180^\circ\) denotes alignment with the nitrogen end pointing toward \(X\).

The radial and angular coordinates are discretized using \(\Delta r=0.1~\text{\AA}\) and \(\Delta\theta=2^\circ\), respectively.
For a bin \([r_m,r_{m+1})\times[\theta_l,\theta_{l+1})\), the SDF is defined as
\begin{equation}
s_X(r_m,t,\theta_l)
=
\frac{
H_X(r_m,t,\theta_l)
}{
N_{\mathrm{frames}}\,
\rho_{\mathrm{ACN}}^{\mathrm{bulk}}\,
\Delta V_{ml}
}.
\end{equation}
Here, \(H_X\) is the count of acetonitrile molecules with distances and orientations falling within the bin, considering all trajectory frames contributing at delay \(t\).
\(N_{\mathrm{frames}}\) is the number of contributing frames, and \(\rho_{\mathrm{ACN}}^{\mathrm{bulk}}\) is the bulk acetonitrile number density.
The effective bin volume is the product of the radial shell volume and the fraction of isotropic molecular orientations within the angular interval:
\begin{equation}
\Delta V_{ml}
=
\left[
\frac{4\pi}{3}
\left(r_{m+1}^{3}-r_m^{3}\right)
\right] \cdot
\left[
\frac{1}{2}
\left(\cos\theta_l-\cos\theta_{l+1}\right)
\right].
\end{equation}
The first factor is the spatial volume of the shell around \(X\), while the second follows from the isotropic angular probability density \(\tfrac12\sin\theta\) and equals unity over the full range \(0\leq\theta\leq\pi\).
Consequently, \(\rho_{\mathrm{ACN}}^{\mathrm{bulk}}\Delta V_{ml}\) is the expected number of molecules per frame in a bin for the isotropic solvent. Thus, for a homogeneous solvent with random molecular orientations, \(s_X(r,t,\theta)=1\).

The SDF describes the joint dependence of the local solvent population on distance and relative molecular orientation.
Radial and angular projections provide complementary measures of the structural redistribution.
The RDF is obtained by averaging over all molecular orientations,
\begin{equation}
\begin{aligned}
g_X(r,t)
&=
\frac{
\displaystyle\int_0^\pi s_X(r,t,\theta)\sin\theta\,d\theta
}{
\displaystyle\int_0^\pi\sin\theta\,d\theta
}\\
&=
\frac{1}{2}
\int_0^\pi
s_X(r,t,\theta)\sin\theta\,d\theta .
\end{aligned}
\end{equation}
The RDF describes the radial distribution of acetonitrile \(N\!-C_{\mathrm{Me}}\) midpoints around \(X\), averaged over molecular orientations.
Integrating the difference RDF over a selected radial interval gives the change in coordination number relative to the equilibrium ground state,
\begin{equation}
\Delta N_X^{\mathrm{RDF}}(t)
=
4\pi\rho_{\mathrm{ACN}}^{\mathrm{bulk}}
\int_{r_1}^{r_2}
\Delta g_X(r,t)\,r^2\,dr ,
\end{equation}
where \(\Delta g_X\) is the difference between the time-dependent and equilibrium ground-state RDFs.
Setting \(r_1=0\) gives the change in the cumulative coordination number within \(r_2\).

The angular distribution function (ADF) is obtained by averaging the SDF over a selected radial shell,
\begin{equation}
\begin{aligned}
a_X(\theta,t;r_1,r_2)
&=
\frac{
\displaystyle\int_{r_1}^{r_2}s_X(r,t,\theta)\,4\pi r^2\,dr
}{
\displaystyle\int_{r_1}^{r_2}4\pi r^2\,dr
}\\
&=
\frac{
\displaystyle
\int_{r_1}^{r_2}
s_X(r,t,\theta)\,r^2\,dr
}{
\displaystyle
\int_{r_1}^{r_2}
r^2\,dr
}.
\end{aligned}
\end{equation}
Restricting the average to a finite shell restricts the analysis on the local orientational preference.
The ADF retains the normalization to the bulk number density, so \(a_X=1\) corresponds to the homogeneous, isotropic bulk solvent.
The change in molecular population within a radial shell and given angular interval is
\begin{equation}
\begin{split}
\Delta N_X^{\mathrm{ADF}}
(t)
&=
\rho_{\mathrm{ACN}}^{\mathrm{bulk}}\frac{4\pi}{3}
\left(r_2^3-r_1^3\right)\\
&\quad\times \frac{1}{2}
\int_{\theta_1}^{\theta_2}
\Delta a_X(\theta,t;r_1,r_2)
\sin\theta\,d\theta ,
\end{split}
\end{equation}
where \(\Delta a_X\) is the difference between the time-dependent and equilibrium ground-state ADFs for the same radial shell.

\subsubsection{Excluded volume}

The volume inaccessible to the solvent surrounding the solute, i.e. the excluded volume, is estimated from a cavity defined by the instantaneous solvent configuration around HTI.
For each trajectory frame, each solvent atom \(j\) is represented by a sphere with radius
\begin{equation}
R_j = r_j^{\mathrm{vdW}} + r_{\mathrm{probe}},
\qquad
r_{\mathrm{probe}} = 1.60~\text{\AA},
\end{equation}
where \(r_j^{\mathrm{vdW}}\) is the van der Waals radius of atom \(j\), taken from Bondi,\cite{Bondi_1964} and \(r_{\mathrm{probe}}\) is a uniform radius increment used to expand the solvent spheres and define the cavity boundary.
A cavity surrounding HTI is identified as the region enclosed by the surface defined by the union of the solvent-centered spheres. The excluded volume is then obtained as the volume of the cavity. The time-dependent excluded volume is obtained by averaging the cavity volumes over trajectories at each delay time. A subset of 5,000 nonequilibrium trajectories was used for this analysis. Frames without a closed cavity surface are rejected and excluded from the average. The chosen probe radius results in the exclusion of approximately \(20\%\) of all frames for a given time delay. 
Further details of the cavity construction and numerical discretization of the volume are provided in the Supporting Information (SI).

\subsubsection{Time-resolved X-ray scattering signal}
Time-resolved scattering signals are calculated from the RDFs between atom types following the formalism described by Dohn \textit{et al.} \cite{Dohn_2016b,Dohn_2015}.
Here, \(q\) denotes the magnitude of the scattering vector (momentum transfer).
The total difference signal is decomposed into solute--solute, solute--solvent, and solvent--solvent contributions,
\begin{equation}
\Delta S_{\mathrm{tot}}(q,t)
=
\Delta S_{\mathrm{uu}}(q,t)
+
\Delta S_{\mathrm{uv}}(q,t)
+
\Delta S_{\mathrm{vv}}(q,t).
\end{equation}
For each scattering component \(c\in\{\mathrm{uu},\mathrm{uv},\mathrm{vv}\}\), the difference signal is evaluated as the scattering from the nonequilibrium distribution at delay \(t\) minus the scattering from the equilibrium ground-state distribution,
\begin{equation}
\Delta S_c(q,t)
=
S_c^{\mathrm{noneq}}(q,t)-S_c^{\mathrm{GS,eq}}(q).
\end{equation}
The difference scattering signal defined in this way simulates the pump-on minus pump-off difference signal measured in TR-XSS experiments.

Before calculating the solute--solvent and solvent--solvent scattering contributions, the corresponding RDFs are corrected for finite-size effects using the Ganguly and van der Vegt correction\cite{Ganguly_2013}. To suppress termination artifacts associated with the finite extent of the RDFs, the corrected RDFs are subsequently multiplied by a Lorch damping window\cite{Lorch_1969}, with the damping length set to the maximum sampled RDF distance of $20 \AA$.\cite{Dohn_2023} The same correction and damping procedure is applied to the nonequilibrium and equilibrium ground-state RDFs. The solute--solute contribution is evaluated directly from the uncorrected RDFs without damping.
The scattering signal is evaluated at equally spaced \(q\) values with a spacing of \(\Delta q=0.01~\text{\AA}^{-1}\).
For the kinetic analysis, selected $q$-regions of the solute--solvent contribution are reduced to one-dimensional time traces by integration over the corresponding \(q\) interval,
\begin{equation}
I(t;q_1,q_2)
=
\int_{q_1}^{q_2}
\Delta S_{\mathrm{uv}}(q,t)\,dq .
\end{equation}
The scattering calculations described above were performed using the Python package \texttt{grsq}.\cite{grsq}

\section{\label{sec:results}Results}
\subsection{Photoinduced charge transfer}
\subsubsection{Local solvent reorganization}
\begin{figure*}[t]
    \centering    
    \includegraphics[width=\textwidth,height=0.38\textheight,keepaspectratio]{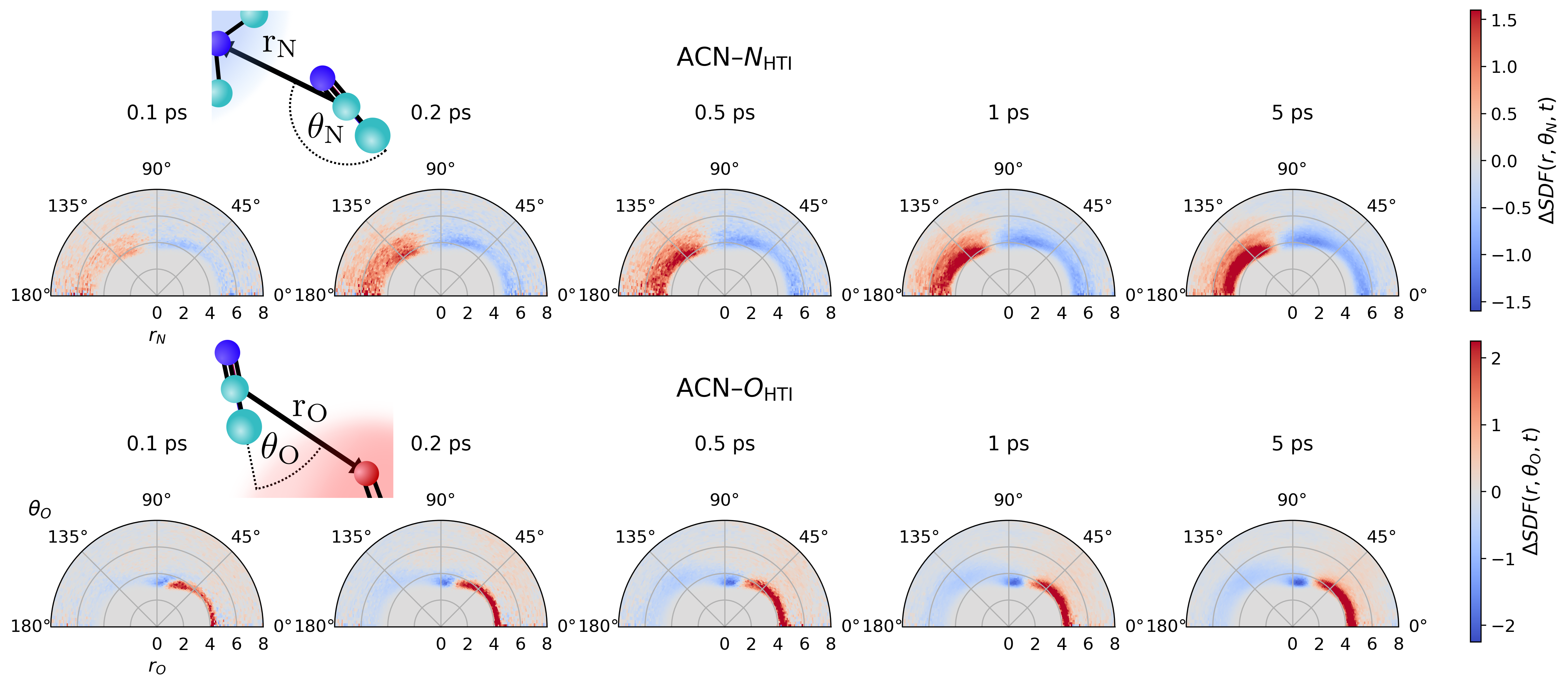}
    \caption{
    Time-resolved difference SDFs of acetonitrile around \(N_{\mathrm{HTI}}\) (top) and \(O_{\mathrm{HTI}}\) (bottom) at selected delay times after charge-transfer excitation, relative to the equilibrium ground-state distributions.
    The radial coordinate \(r_X\) measures the distance from the midpoint between the acetonitrile nitrogen and methyl carbon to the solute atom \(X\), while \(\theta_X\) is the angle between the acetonitrile \(N\!\rightarrow C_{\mathrm{Me}}\) axis and the vector from this midpoint toward \(X\) (Fig.~\ref{fig:SDF_definition}).
    The distributions show a fast initial reorientation, with the nitrile nitrogen pointing toward \(N_{\mathrm{HTI}}\) and the methyl end toward \(O_{\mathrm{HTI}}\), while slower orientational and radial rearrangements continue over the following picoseconds.
    }
    \label{fig:SDFs}
\end{figure*}

Fig.~\ref{fig:SDFs} shows the time-dependent difference SDFs of acetonitrile distances and orientations around \(N_{\mathrm{HTI}}\) and \(O_{\mathrm{HTI}}\) from the nonequilibrium simulations of photoinduced charge transfer, evaluated relative to the equilibrium distributions in the ground state.
The radial coordinate \(r_X\) measures the distance from the acetonitrile \(N\!-C_{\mathrm{Me}}\) midpoint to the solute atom \(X\), while \(\theta_X\) is the angle between the acetonitrile \(N\!\rightarrow C_{\mathrm{Me}}\) axis and the vector from the midpoint toward \(X\) (Fig.~\ref{fig:SDF_definition}).
Around \(N_{\mathrm{HTI}}\), solvent  population is depleted at small and intermediate angles and accumulates at larger \(\theta_N\), corresponding to orientations with the nitrile nitrogen pointing toward the solute nitrogen.
Around \(O_{\mathrm{HTI}}\), population instead increases at small \(\theta_O\), with the methyl end pointing toward the solute oxygen.
The angular redistribution is already apparent at \(0.1\)--\(0.2~\mathrm{ps}\), while both orientational and radial distributions continue to evolve over the following picoseconds.

These opposite orientational responses are consistent with the direction of photoinduced charge transfer illustrated in Fig.~\ref{fig:SDF_definition}. Electron density moves from the julolidine donor region containing \(N_{\mathrm{HTI}}\) toward the thioindigo acceptor region containing \(O_{\mathrm{HTI}}\).
In the force-field model, the corresponding changes in atomic partial charges are \(+0.38\,e\) for the nitrogen and \(-0.21\,e\) for the oxygen, explaining the enhanced preference for the negatively polarized nitrile of ACN end near the former and the positively polarized methyl of ACN end near the latter.

\begin{figure*}[t]
    \centering
    \includegraphics[width=\textwidth,height=0.38\textheight,keepaspectratio]{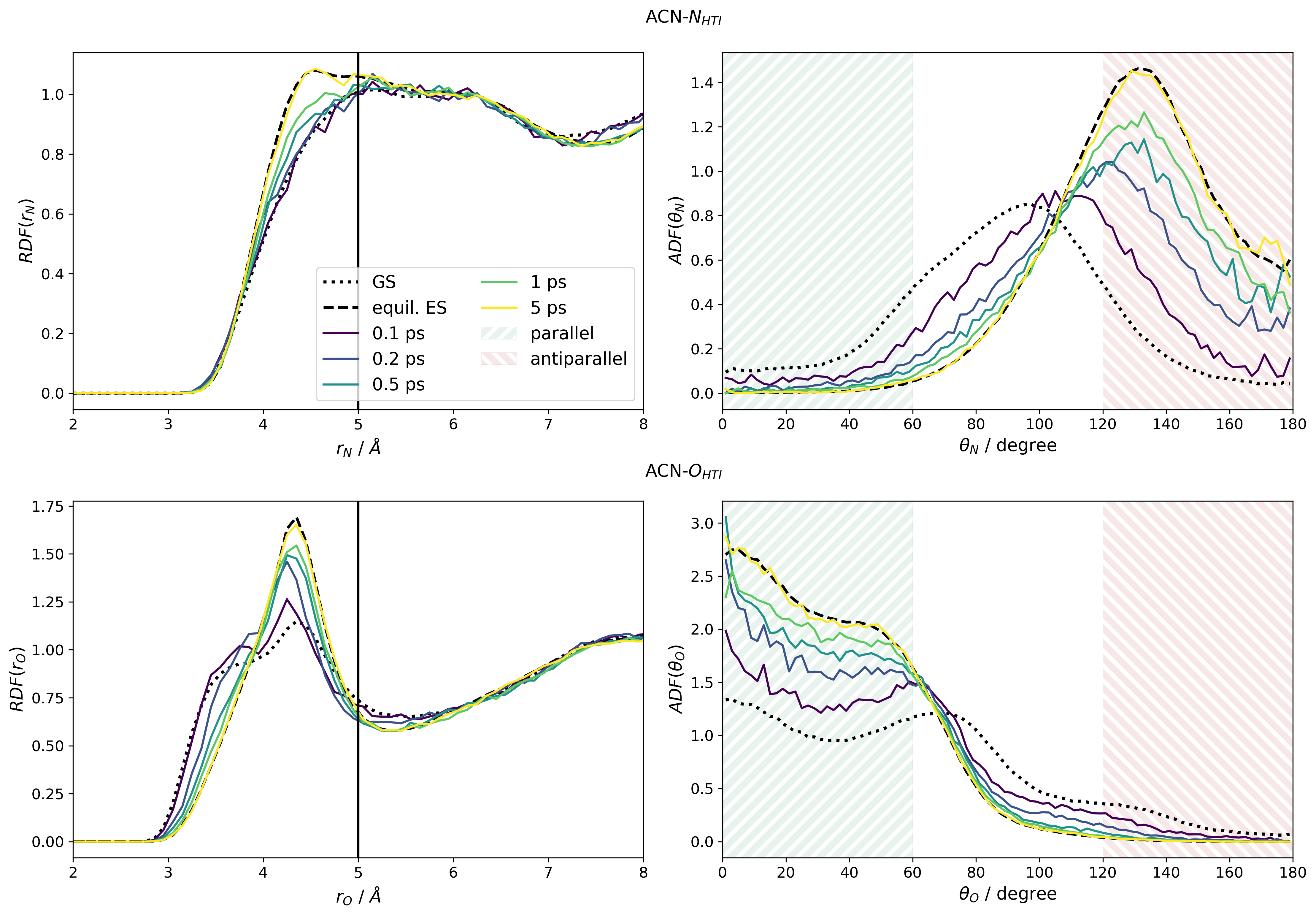}
    \caption{
    Radial and angular distribution functions around \(N_{\mathrm{HTI}}\) (top) and \(O_{\mathrm{HTI}}\) (bottom) at selected delay times after photoinduced charge transfer.
    Left: Radial distribution functions, \(g_X(r,t)\), for the distance \(r\) between the solute atom \(X\) and the midpoint between the nitrogen and methyl carbon of an acetonitrile molecule.
    Right: Angular distribution functions, \(a_X(\theta,t)\), obtained by averaging the spatial distribution functions up to \(r=5~\text{\AA}\) (vertical lines in the left panels) with \(\theta\) being the angle between the acetonitrile \(N\!\rightarrow C_{\mathrm{Me}}\) axis and the vector from its \(N\!-C_{\mathrm{Me}}\) midpoint toward \(X\).
    The hatched areas define parallel (\(0^\circ\leq\theta\leq60^\circ\)) and antiparallel (\(120^\circ\leq\theta\leq180^\circ\)) orientation ranges.
    At early delay times, the angular distributions change more rapidly than the radial distributions, consistent with fast initial solvent reorientation followed by slower rearrangement involving both rotation and translation of solvent molecules.
    }
    \label{fig:SDF_projections}
\end{figure*}

To quantify the molecular reorientation and translational redistribution, we analyze the radial, \(g_X(r,t)\), and angular distribution functions, \(a_X(\theta,t)\) (Fig.~\ref{fig:SDF_projections}).
For \(N_{\mathrm{HTI}}\), the radial and angular distributions evolve on distinct time scales (Fig.~\ref{fig:SDF_projections}, top).
Within the first few hundred femtoseconds, the acetonitrile population shifts from intermediate angles toward larger \(\theta_N\).
Over the following picoseconds, the angular distribution approaches the pronounced maximum at \(130^\circ\)--\(140^\circ\) in the equilibrium excited-state distribution obtained from an equilibrated excited-state trajectory of \(\sim 400~\si{ns}\).
The radial redistribution is slower, as becomes more apparent from the quantitative analysis presented below in Fig.~\ref{fig:population_dynamics}. Although the position of the main first-shell feature changes only slightly, a broad peak develops at shorter distances, becoming clearly apparent between approximately \(1\) and \(5~\mathrm{ps}\) as the RDF approaches the excited-state equilibrium distribution.
The strong initial angular redistribution, accompanied by smaller radial changes, is consistent with reorientation of acetonitrile molecules already present in the local solvation shell, while the subsequent growth of the RDF peak indicates additional translational restructuring on a slower time scale.

The distributions for \(O_{\mathrm{HTI}}\) show a similar temporal hierarchy (Fig.~\ref{fig:SDF_projections}, bottom).
Here, the ground-state angular distribution already favors small \(\theta_O\), with the methyl end pointing toward the oxygen site, consistent with its negative partial charge of approximately \(-0.62\,e\) in the ground state.
After excitation, this orientational preference strengthens rapidly, visible through an increase in population at small angles and depletion at intermediate and large angles.
The first-shell peak in the radial distribution grows more slowly and approaches the excited-state equilibrium distribution between \(1\) and \(5~\mathrm{ps}\).
Both sites thus exhibit rapid molecular reorientation accompanied by slower radial restructuring, despite their different ground-state distributions and opposite preferred orientations.

The time scales of the solvent motions in the first solvation shell are quantified using the time-dependent changes in molecular populations shown in Fig.~\ref{fig:population_dynamics}.
The radial population changes, \(\Delta N_X^{\mathrm{RDF}}\), are evaluated over \(0\leq r_X\leq5~\text{\AA}\), as marked by the vertical lines in Fig.~\ref{fig:SDF_projections}.
This cutoff is motivated by the approximate extent of the first solvation shell in the distribution of distances from the center of mass of each acetonitrile molecule to the closest HTI atom (Fig.~\ref{fig:delocalized_solvent}).
Within the same radial interval, the angular population changes, \(\Delta N_X^{\mathrm{ADF}}\), are evaluated over a range of antiparallel orientations \(120^\circ\leq\theta_N\leq180^\circ\) for \(N_{\mathrm{HTI}}\) and a range of parallel orientations \(0^\circ\leq\theta_O\leq60^\circ\) for \(O_{\mathrm{HTI}}\), defined by the hatched areas in Fig.~\ref{fig:SDF_projections}.
\begin{figure*}[t]
    \centering
    \includegraphics[width=\textwidth,height=0.38\textheight,keepaspectratio]{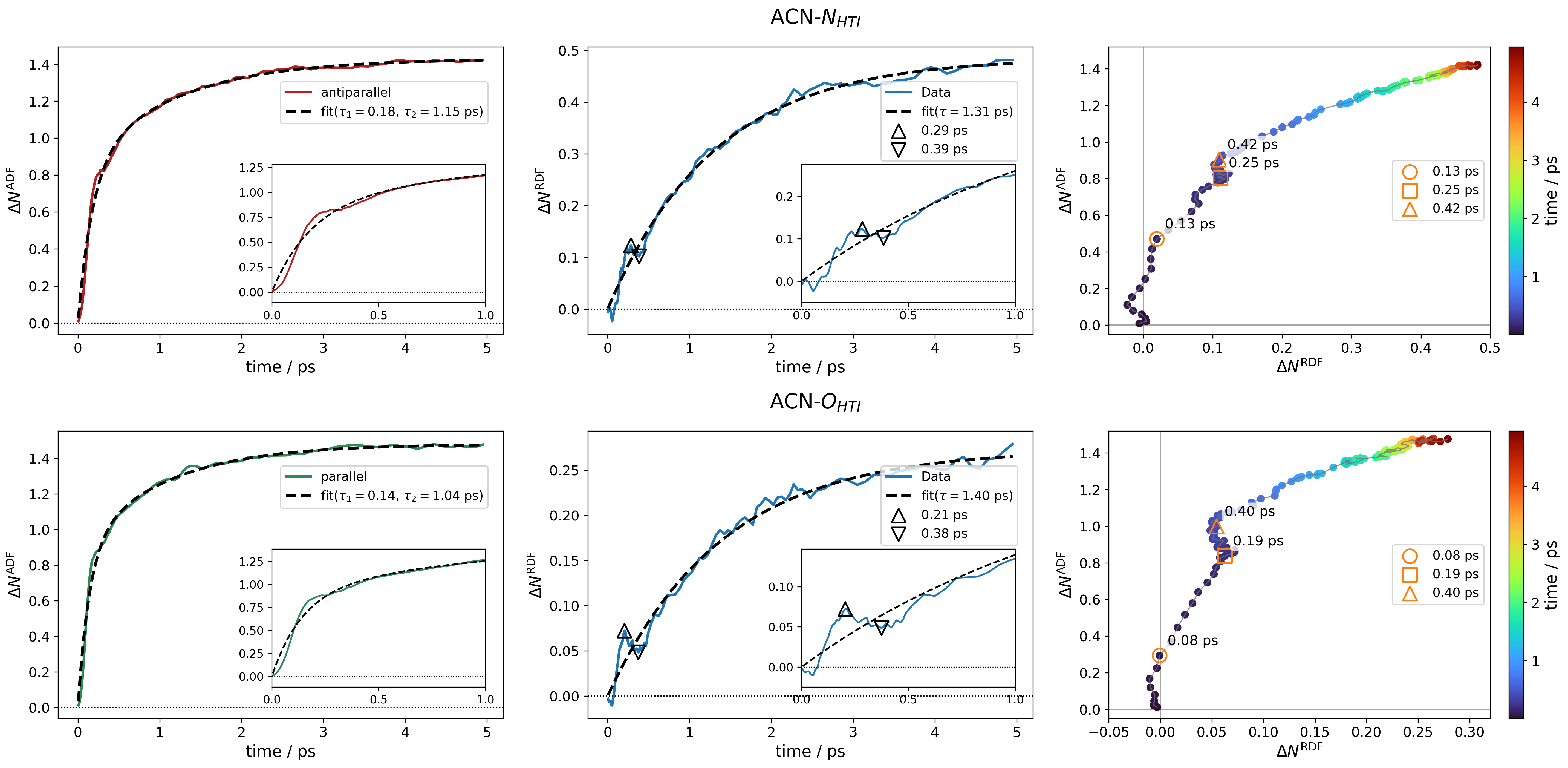}
    \caption{
    Time-dependent radial and orientational population changes around \(N_{\mathrm{HTI}}\) (top) and \(O_{\mathrm{HTI}}\) (bottom), relative to equilibrium in the ground state.
    Left: Orientational population change, \(\Delta N_X^{\mathrm{ADF}}\), obtained by integrating the difference ADFs over \(0\leq r_X\leq5~\text{\AA}\) as well as the antiparallel range for \(N_{\mathrm{HTI}}\) and the parallel range for \(O_{\mathrm{HTI}}\), defined by the hatched areas in Fig.~\ref{fig:SDF_projections}.
    Center: Cumulative coordination number, \(\Delta N_X^{\mathrm{RDF}}\), for \(r_X = 5~\text{\AA}\). The insets highlight the first picosecond.
    Right: Orientational population changes \(\Delta N_X^{\mathrm{ADF}}\) plotted against radial population changes \(\Delta N_X^{\mathrm{RDF}}\), with color showing the delay time after excitation. The marked early-time points highlight initial molecular reorientation with little net change in radial population, followed by slower radial redistribution. 
    The initial dynamics are nonexponential, showing a rapid population rise followed by a brief partial decay at approximately \(0.2\)--\(0.4~\mathrm{ps}\), with the fast solvent response dominated by molecular reorientation.
    }
    \label{fig:population_dynamics}
\end{figure*}
Biexponential fits to the time evolution of \(\Delta N_X^{\mathrm{ADF}}\) yield fast and slow time constants of \(0.18\) and \(1.15~\mathrm{ps}\) for \(N_{\mathrm{HTI}}\), and \(0.14\) and \(1.04~\mathrm{ps}\) for \(O_{\mathrm{HTI}}\).
These fits yield average relaxation times of approximately \(\mathrm{0.58}\) and \(\mathrm{0.50}~\mathrm{ps}\) for \(N_{\mathrm{HTI}}\) and \(O_{\mathrm{HTI}}\), respectively, close to the \(0.3\pm 0.1~\mathrm{ps}\) time scale previously extracted from the HTI TR-XSS data for a component attributed to solvation-shell rearrangement.\cite{Mitterer_2026}
The radial populations evolve more slowly, with fitted time constants of \(1.31\) and \(1.40~\mathrm{ps}\), respectively. However, the early response departs from an exponential behavior. Both \(\Delta N_X^{\mathrm{ADF}}\) and \(\Delta N_X^{\mathrm{RDF}}\) rise rapidly with a nonexponential onset, followed by a partial reversal at approximately \(0.2\)--\(0.4~\mathrm{ps}\), before the  rise resumes.
The fitted time constants therefore characterize the relaxation without fully describing this initial nonexponential dynamics.

The right panels of Fig.~\ref{fig:population_dynamics} make the temporal ordering of rotations and translations more clear by plotting \(\Delta N_X^{\mathrm{ADF}}\) against \(\Delta N_X^{\mathrm{RDF}}\).
During approximately the first \(100\)--\(150~\mathrm{fs}\), the orientational population increases while the radial population remains close to its ground-state value.
This indicates predominantly molecular reorientation with little net change of the number of molecules in the first solvation shell. Radial redistribution follows and evolves more slowly.
The similar time scales at the two sites suggest a common solvent relaxation, despite their opposite preferred orientations.\\
An analogous analysis of the SDFs around the sulfur atom \(S_{\mathrm{HTI}}\) is reported in the SI.

\subsubsection{Excluded volume}

\begin{figure}[t]
    \centering
    \includegraphics[width=\columnwidth]{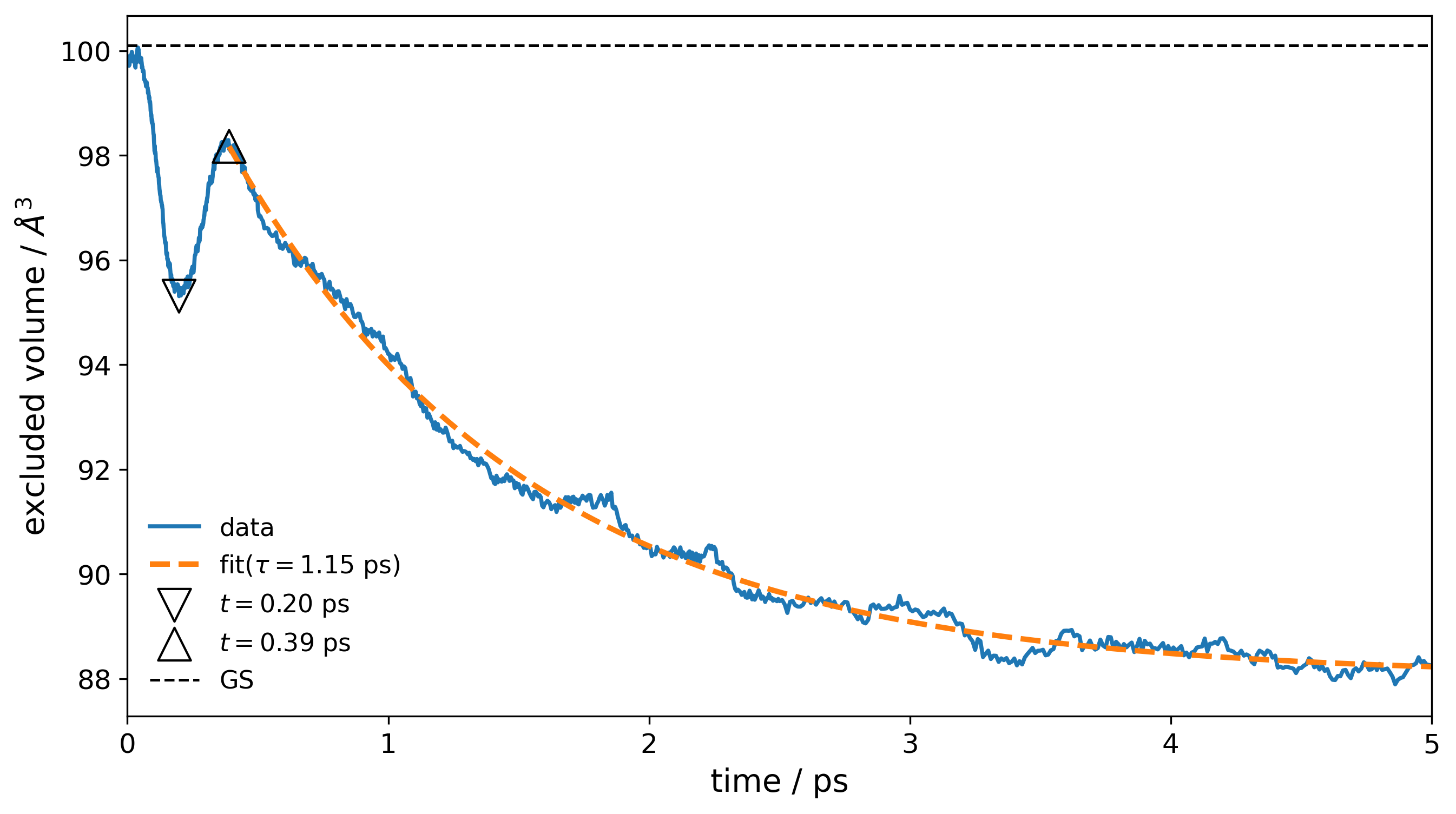}
    \caption{
    Time dependence of the volume around the HTI inaccessible to the solvent (excluded volume) following charge-transfer excitation.
    The dashed horizontal line marks the equilibrium ground-state value, \(V_{\mathrm{EV}}^{\mathrm{GS}}=100.1~\text{\AA}^3\).
    Triangles mark the minimum and the subsequent local maximum in the time evolution.
    The excluded volume shows a fast nonexponential initial contraction, followed by a transient rebound and slower relaxation on the picosecond time scale.
    }
    \label{fig:excluded_volume}
\end{figure}

The preceding analysis reveales a rapid reorientation of acetonitrile molecules within the first solvation shell after excitation. Here, we examine whether this fast orientational response is accompanied by a contraction of the volume inaccessible to the solvent around the solute (excluded volume).
The associated changes in solvent packing around the solute can in turn induce rearrangements in outer solvation shells and changes in local solvent density.
The resulting changes in intermolecular distances are expected to affect both the solute--solvent and solvent--solvent scattering contributions.

Figure~\ref{fig:excluded_volume} shows the excluded volume as a function of time after charge-transfer excitation.
The equilibrium ground-state value is \(V_{\mathrm{EV}}^{\mathrm{GS}}=100.1~\text{\AA}^3\).
Initially, the volume contracts to a local minimum of \(95.3~\text{\AA}^3\) at \(\sim 200~\mathrm{fs}\), then rebounds to a local maximum of \(98.2~\text{\AA}^3\) at \(\sim400~\mathrm{fs}\).
From approximately \(400~\mathrm{fs}\) onward, a slower contraction follows, described by a single-exponential decay with a time constant of \(1.15~\mathrm{ps}\).
At \(5~\mathrm{ps}\), the excluded volume reaches \(88.2~\text{\AA}^3\), corresponding to a net contraction of approximately \(12~\text{\AA}^3\), or \(\sim12\%\).

This evolution is consistent with the angular population dynamics in Fig.~\ref{fig:population_dynamics}.
The initial contraction occurs on a time scale comparable to the fast orientational components of \(0.15\)--\(0.2~\mathrm{ps}\), while the subsequent contraction has a time constant close to the slower orientational components of \(1.04\)--\(1.15~\mathrm{ps}\).
Reorientation changes the positions of the solvent atoms defining the cavity boundary and can therefore reduce the excluded volume, while the number of molecules in the first solvation shell remains nearly constant.
The excluded volume thus captures changes in the local geometry of the solvent shell surrounding HTI.

The initial nonexponential contraction and subsequent rebound suggest a transient readjustment of the first solvation shell following its rapid reorientation. 
A possible interpretation is that rapid reorientation driven by the torque due to the changed electrostatic field of the solute brings solvent atoms into closer contact with the solute and with each other. This increases the short-range repulsive interactions that oppose the initial motion.
The resulting transient outward readjustment produces a partial expansion of the shell before slower relaxation continues.
The time evolution of the excluded volume highlights this rebound more clearly than the smaller partial reversals observed in the local population curves (Fig.~\ref{fig:population_dynamics}).
This is because it reflects the combined geometric response of the whole solvation shell, while the population curves probe local rearrangements relative to individual sites.

\subsubsection{Solvent redistribution across solvation shells}

\begin{figure*}[t]
    \centering
    \includegraphics[width=0.8\textwidth]{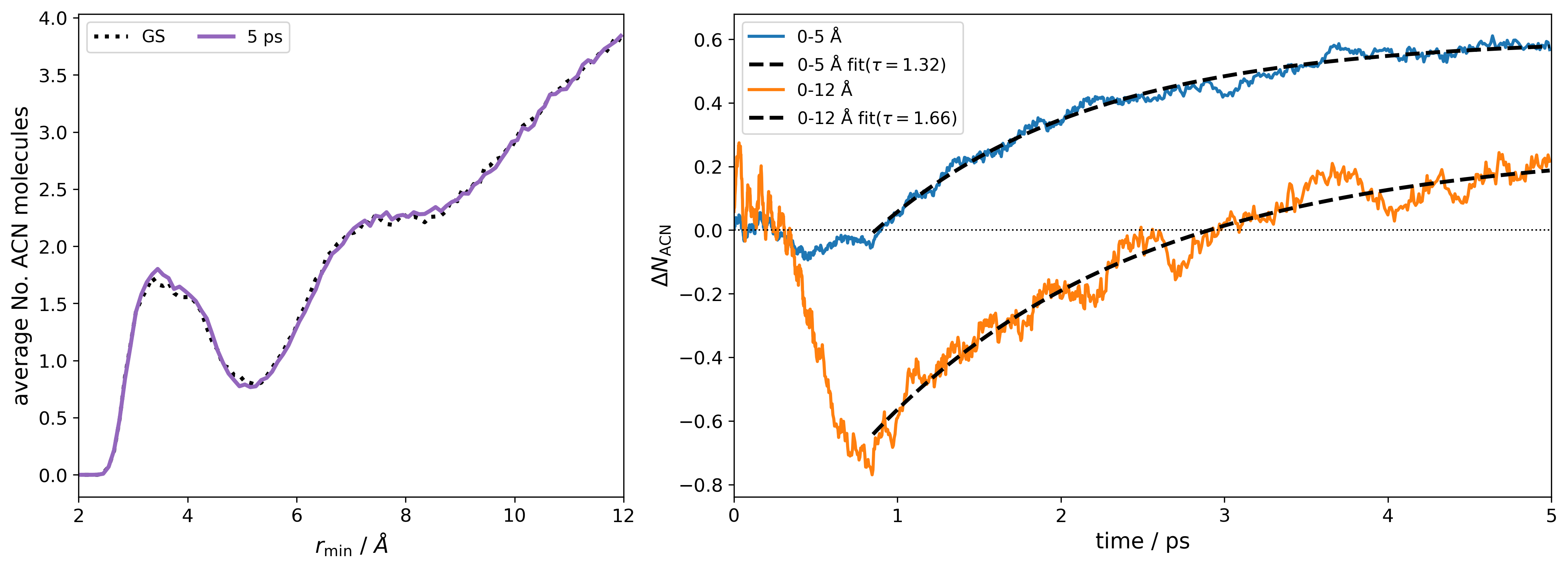}
    \caption{
    Left: Ensemble-averaged number of acetonitrile molecules per \(r_{\min}\) bin (bin width: \(0.1~\text{\AA}\)), where \(r_{\min}\) is the distance from the center of mass of each molecule to the closest \mbox{HTI} atom, shown for the equilibrium ground state (dotted line) and at \(t=5~\mathrm{ps}\) after charge-transfer excitation (solid line).
    Right: Change in the cumulative number of acetonitrile molecules, obtained by integrating the number distribution up to \(r_{\min}=5~\text{\AA}\) (approximately the first solvation shell) and \(r_{\min}=12~\text{\AA}\) (including outer shells), relative to the equilibrium ground-state population.
    The population of the first shell increases after about \(0.9~\mathrm{ps}\), while the cumulative population within \(12~\text{\AA}\) shows initial depletion up to \(\sim0.9~\mathrm{ps}\) followed by recovery on a longer time scale.
    }
    \label{fig:delocalized_solvent}
\end{figure*}

The preceding analyses describe solvent rearrangements relative to selected HTI atom sites and are largely restricted to the first solvation shell.
To examine the redistribution within the first and outer solvation shells, we consider the distance \(r_{\min}\) from the center of mass of each acetonitrile molecule to the closest HTI atom. For this analysis the same subset of 5,000 nonequilibrium trajectories as in the excluded volume analysis was used.
The left panel of Fig.~\ref{fig:delocalized_solvent} shows the ensemble-averaged distribution of these distances in the equilibrium ground state and at \(5~\mathrm{ps}\). The right panel shows the change in the cumulative number of acetonitrile molecules, obtained by integrating the corresponding distribution up to \(r_{\min}=5~\text{\AA}\) or \(r_{\min}=12~\text{\AA}\) and subtracting the equilibrium ground-state value.
The smaller interval approximately encompasses the first solvation shell, while the larger interval also includes outer shells.

The first-shell population remains approximately constant until \(\sim0.9~\mathrm{ps}\), then increases with a fitted time constant of \(1.32~\mathrm{ps}\), reaching a mean net gain of approximately \(0.6\) molecules at \(5~\mathrm{ps}\).
This time constant is consistent with the \(1.31\)--\(1.40~\mathrm{ps}\) relaxation obtained from the local RDF populations (Fig.~\ref{fig:population_dynamics}).
Within \(12~\text{\AA}\), the population instead undergoes significant initial depletion, reaching a mean loss of approximately \(0.7\) molecules at \(\sim0.9~\mathrm{ps}\).
A slower recovery follows, with a fitted time constant of \(1.66~\mathrm{ps}\).
The cumulative population returns to its ground-state value at approximately \(3~\mathrm{ps}\) and subsequently increases, reaching a mean net gain of approximately \(0.2\) molecules at \(5~\mathrm{ps}\).

These observations are consistent with a transient breathing response of the solvation structure.
The transient expansion of the excluded volume identified in Fig.~\ref{fig:excluded_volume} may induce outward displacement of surrounding solvent layers, temporarily reducing the number of molecules within \(12~\text{\AA}\).
Slower inward relaxation then restores this population.
At \(t=5~\mathrm{ps}\), the first-shell gain exceeds the net gain within \(12~\text{\AA}\), indicating that the \(5\)--\(12~\text{\AA}\) region remains depleted compared to the ground state although overall up to 12 Å, the number of solvent molecules increases.

\subsubsection{Time-resolved X-ray scattering signal}

\begin{figure*}[t]
    \centering

    \begin{subfigure}[t]{0.49\textwidth}
        \centering
        \includegraphics[width=\linewidth]{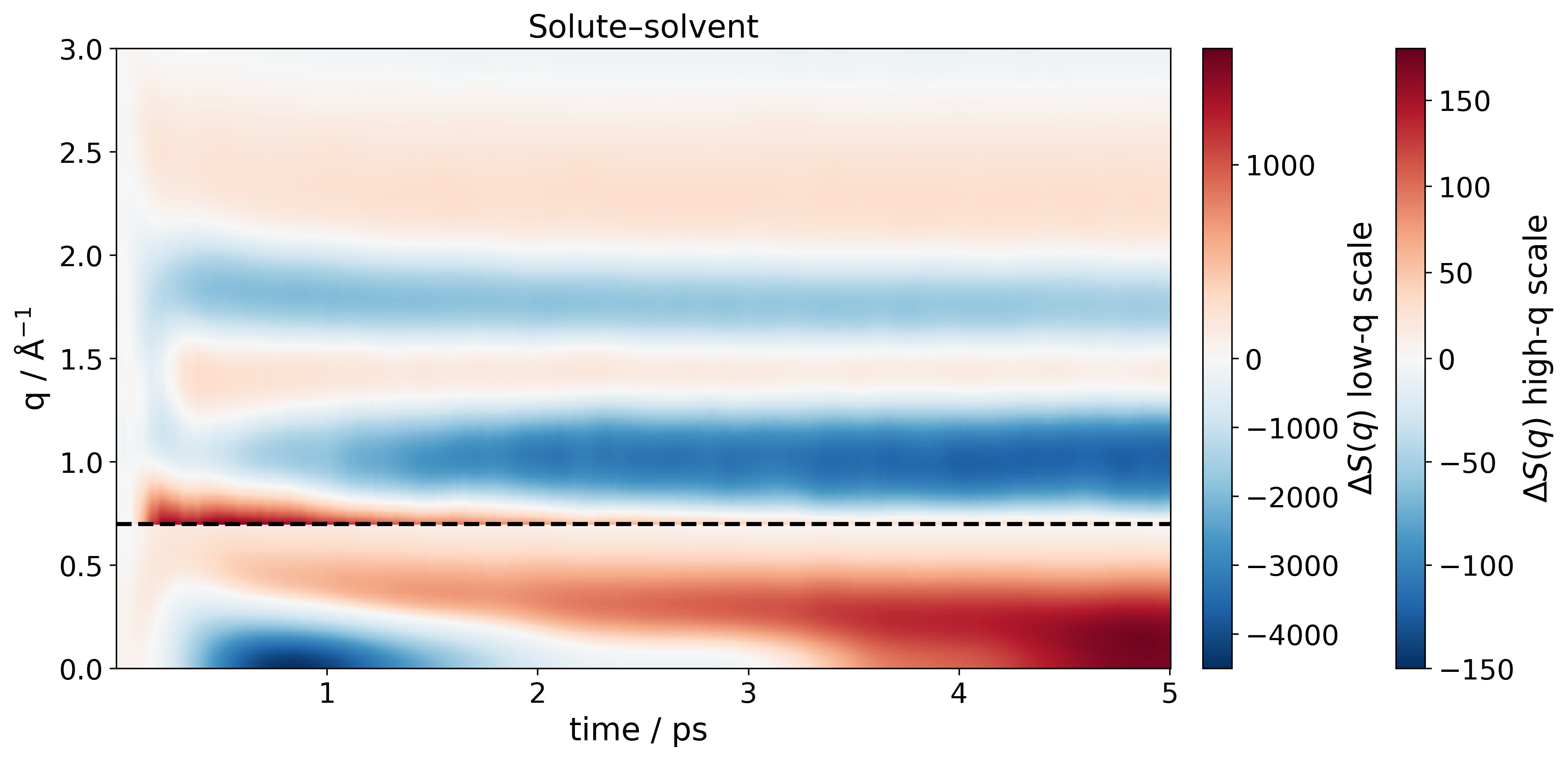}
        \caption{}
        \label{fig:scattering_uv}
    \end{subfigure}
    \hfill
    \begin{subfigure}[t]{0.49\textwidth}
        \centering
        \includegraphics[width=\linewidth]{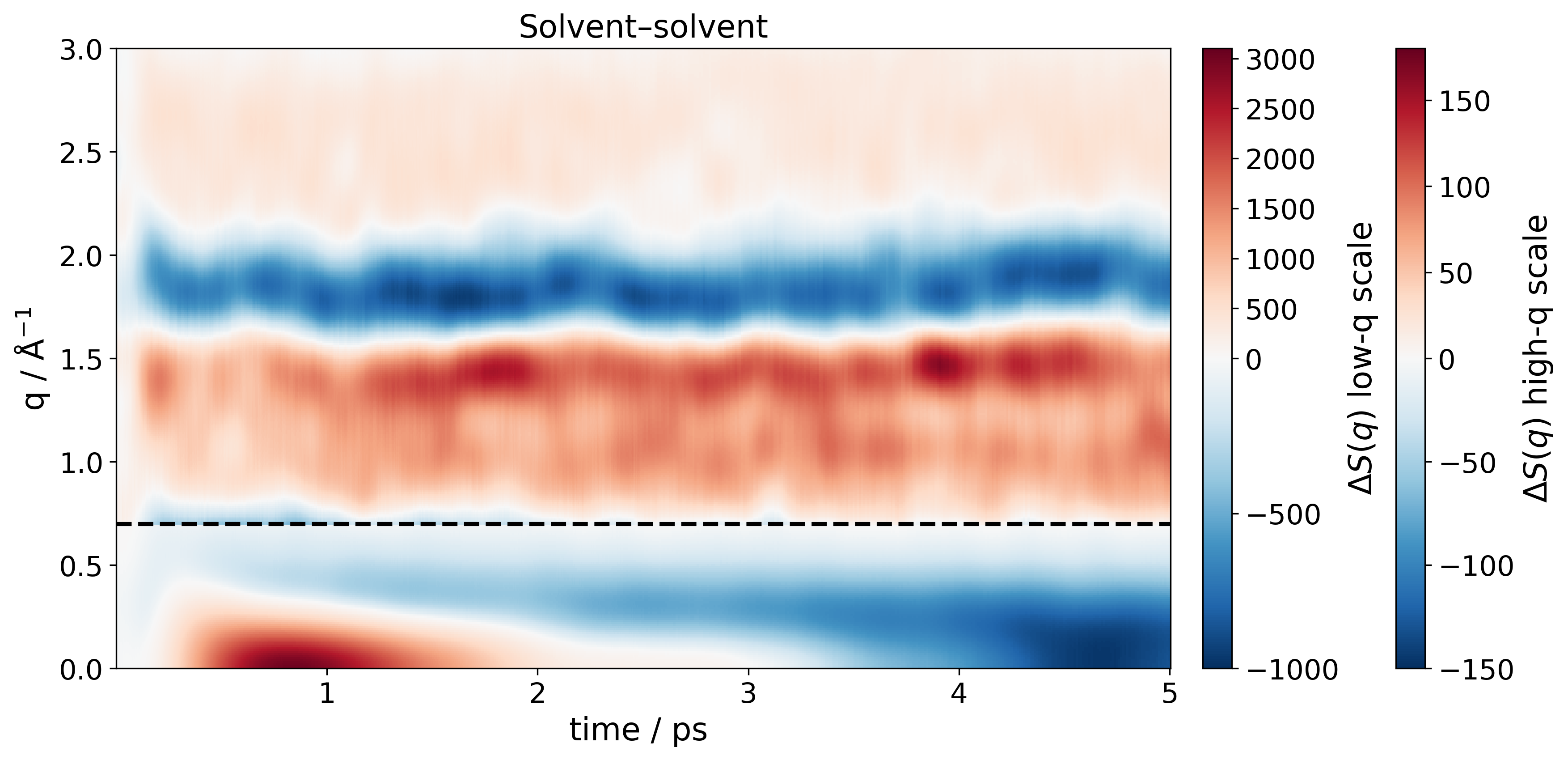}
        \caption{}
        \label{fig:scattering_vv}
    \end{subfigure}

    \caption{
    Time-resolved difference X-ray scattering signals following charge-transfer excitation, calculated relative to the equilibrium ground-state scattering as functions of the magnitude of the scattering vector, \(q\), and time.
    Left: Solute--solvent contribution, \(\Delta S_{\mathrm{uv}}(q,t)\).
    Right: Solvent--solvent contribution, \(\Delta S_{\mathrm{vv}}(q,t)\).
    The dashed horizontal line at \(q=0.7~\text{\AA}^{-1}\) separates the low- and higher-\(q\) regions, displayed using separate color scales.
    Low-\(q\) features reflect both changes in local solvent packing as well as larger-amplitude redistribution across solvation shells. Higher-\(q\) solute--solvent features primarily reflect rapid reorientation, rebound, and slower relaxation in the first solvation shell, while higher-\(q\) solvent--solvent features show the characteristic solvent-heating response.
    }
    \label{fig:scattering_components}
\end{figure*}

\begin{figure}[t]
    \centering
    \includegraphics[width=\columnwidth]{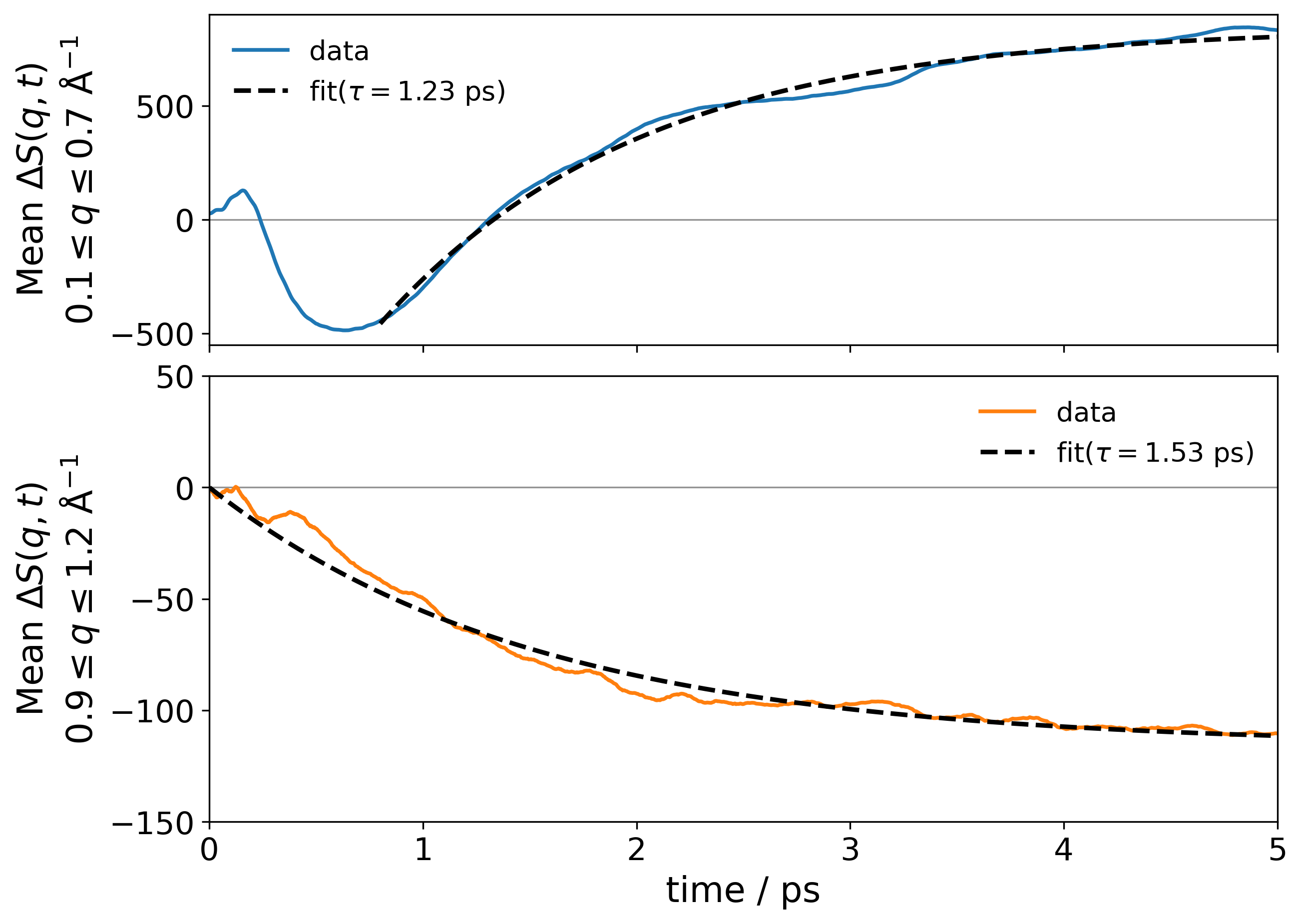}
    \caption{
    Time evolution of the solute--solvent difference X-ray scattering signal in Fig.~\ref{fig:scattering_components} over selected $q$ ranges, following charge-transfer excitation.
    Top: \(\Delta S_{\mathrm{uv}}(q,t)\) averaged over \(0.1\leq q\leq0.7~\text{\AA}^{-1}\).
    Bottom: The same contribution averaged over \(0.9\leq q\leq1.2~\text{\AA}^{-1}\).
    }
    \label{fig:scattering_kinetics}
\end{figure}

Figure~\ref{fig:scattering_components} shows the simulated time-resolved solute--solvent and solvent--solvent difference scattering contributions, \(\Delta S_{\mathrm{uv}}(q,t)\) and \(\Delta S_{\mathrm{vv}}(q,t)\), following charge-transfer excitation, relative to the equilibrium ground-state scattering, as a function of the magnitude of the scattering vector, \(q\), and time.
The strongest amplitudes occur in the low-\(q\) region, \(q\leq0.7~\text{\AA}^{-1}\), where the two contributions exhibit similar features with opposite sign.

Figure~\ref{fig:scattering_kinetics} shows the time evolution of the solute--solvent difference signal averaged over two \(q\) intervals, \(0.1\leq q\leq0.7~\text{\AA}^{-1}\) and \(0.9\leq q\leq1.2~\text{\AA}^{-1}\), together with monoexponential fits.
In the low-\(q\) interval, a small initial positive signal is followed by a negative minimum at \(\sim 0.7~\mathrm{ps}\).
The signal then recovers and rises to positive values, with a fitted relaxation time of \(1.23~\mathrm{ps}\).
This evolution resembles the initial depletion and subsequent recovery of the cumulative solvent population within \(12~\text{\AA}\) (Fig.~\ref{fig:delocalized_solvent}), while the fitted time constant is close to the \(1.32~\mathrm{ps}\) increase in the population of the first shell.
These similarities suggest that the low-\(q\) signal reflects redistribution within and across the solvation shells.
Sec.~IV of the SI shows a scattering-like transform of the change in number of solvent molecules obtained from the minimum distance distribution, which provides further support for this interpretation.

In the \(0.9\)--\(1.2~\text{\AA}^{-1}\) interval (Fig.~\ref{fig:scattering_kinetics}, bottom), the absolute amplitude of the negative signal initially increases rapidly, then decreases briefly before growing on a slower time scale.
The monoexponential fit gives a relaxation time of \(1.53~\mathrm{ps}\) and describes the slower evolution, but does not capture the initial nonexponential dynamics and small rebound within the first \(0.4~\mathrm{ps}\).
These features parallel the rapid reorientation, transient rebound, and slower relaxation of the local first solvation shell identified in the angular populations and excluded volume (Figs.~\ref{fig:population_dynamics} and \ref{fig:excluded_volume}).
The higher-\(q\) solvent--solvent contribution instead displays the characteristic solvent-heating pattern: a positive band around \(1.3\)--\(1.5~\text{\AA}^{-1}\), a negative band near \(1.8\)--\(2.0~\text{\AA}^{-1}\), and a weaker positive feature around \(2.4~\text{\AA}^{-1}\) (Fig.~\ref{fig:scattering_components}, right).
This pattern is consistent with the scattering changes caused by a temperature increase in acetonitrile at constant volume, reflecting changes in solvent intermolecular distances.\cite{Mitterer_2026,Kjaer_2013}

\subsection{Vibrational excitation}
\begin{figure*}[t]
    \centering
    \includegraphics[width=0.8\textwidth]{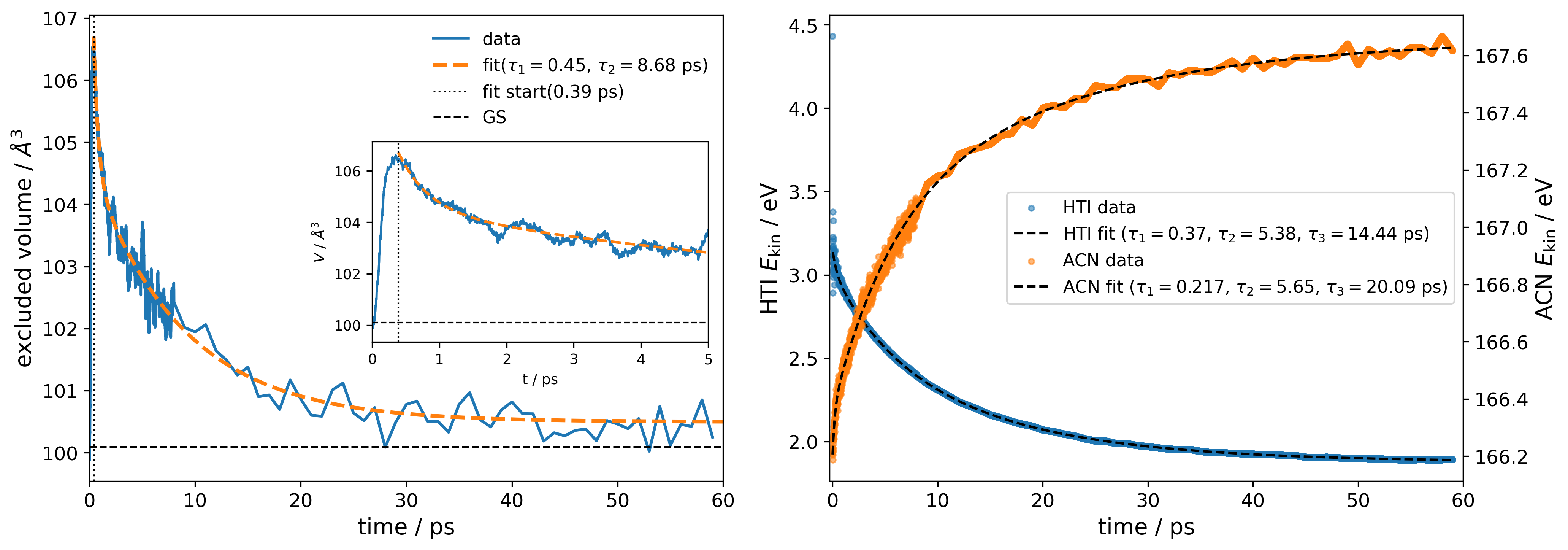}
    \caption{
    Time-dependent excluded volume and kinetic energy following vibrational excitation of \mbox{HTI} in acetonitrile.
    Left: volume around \mbox{HTI} inaccessible to the solvent (excluded volume), with the inset showing the first \(5~\mathrm{ps}\).
    The horizontal dashed line marks the equilibrium ground-state value, \(V_{\mathrm{EV}}^{\mathrm{GS}}=100.1~\text{\AA}^{3}\).
    Right: kinetic energy of \mbox{HTI} (left axis) and acetonitrile (right axis).
    Following the initial drop in solute kinetic energy, the excluded volume reaches a maximum at \(\sim0.4~\mathrm{ps}\) and then contracts as the solute cools and the solvent gains kinetic energy.
    }  \label{fig:heating_cavity_energy}
\end{figure*}

\begin{figure*}[t]
    \centering

    \begin{subfigure}[t]{0.49\textwidth}
        \centering
        \includegraphics[width=\linewidth]{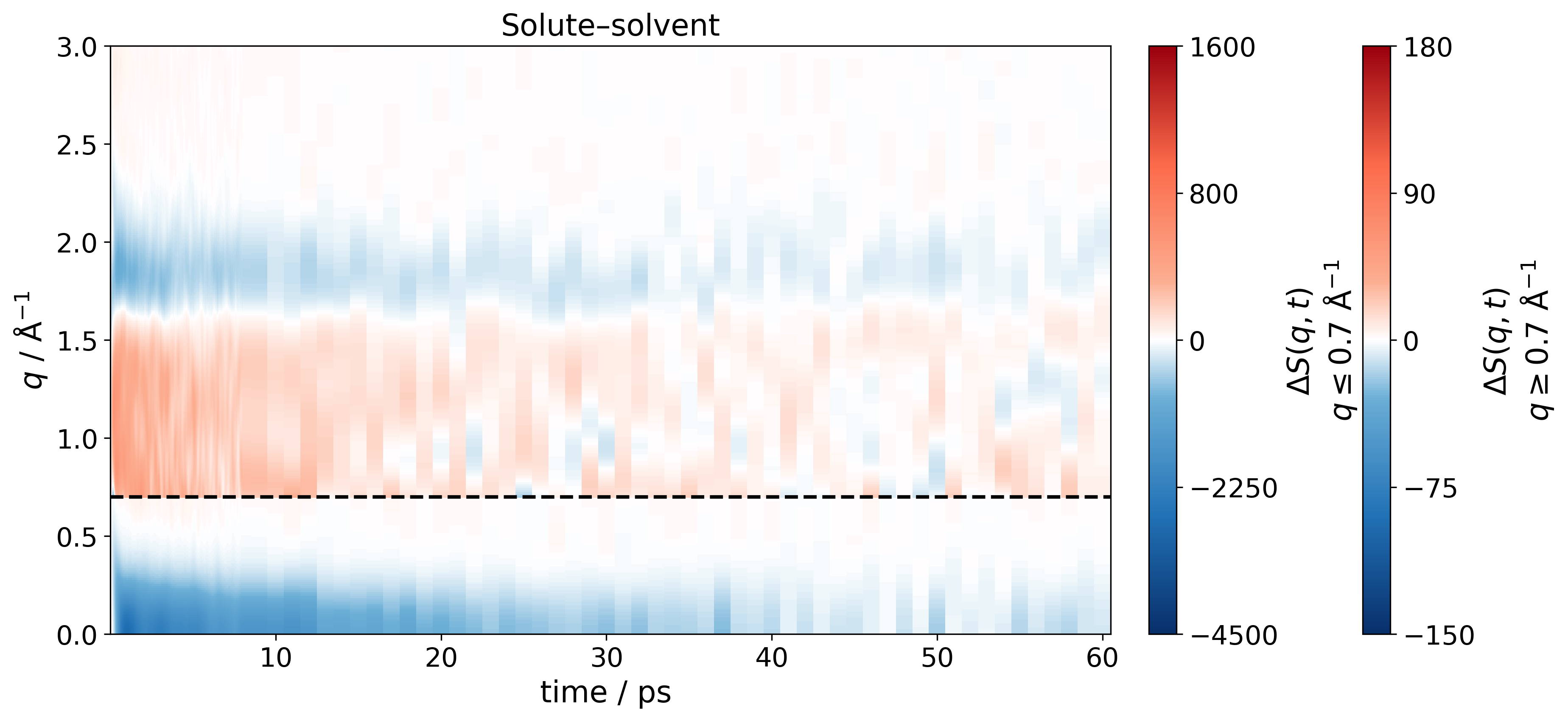}
        \caption{}
        \label{fig:heating_scattering_uv}
    \end{subfigure}
    \hfill
    \begin{subfigure}[t]{0.49\textwidth}
        \centering
        \includegraphics[width=\linewidth]{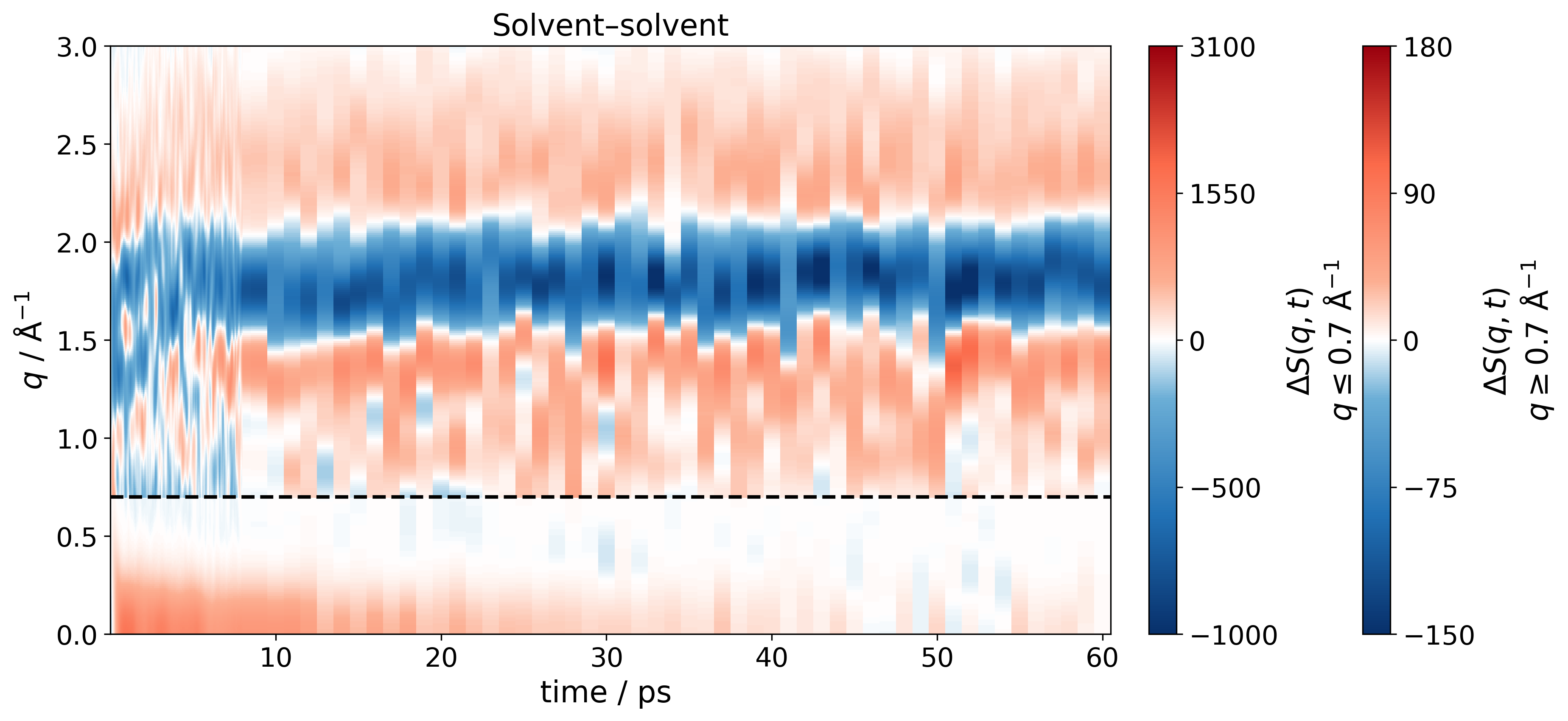}
        \caption{}
        \label{fig:heating_scattering_vv}
    \end{subfigure}

    \caption{
    Time-resolved difference X-ray scattering signals following vibrational excitation of \mbox{HTI}, calculated relative to the equilibrium ground-
     state as a function of the magnitude of the scattering vector, \(q\), and time.
    Left: Solute--solvent contribution, \(\Delta S_{\mathrm{uv}}(q,t)\).
    Right: Solvent--solvent contribution, \(\Delta S_{\mathrm{vv}}(q,t)\).
    The dashed horizontal line at \(q=0.7~\text{\AA}^{-1}\) separates the low- and higher-\(q\) regions, displayed using separate color scales.
    Low-\(q\) and higher-\(q\) features in the solute--solvent signal are associated with restructuring of the solvation shell, with the high-q features more closely reflecting the time evolution of the excluded volume, while higher-\(q\) solvent--solvent features show the bulk-solvent heating response.
    }
    \label{fig:heating_scattering}
\end{figure*}

\begin{figure}[t]
    \centering
    \includegraphics[width=0.9\columnwidth]{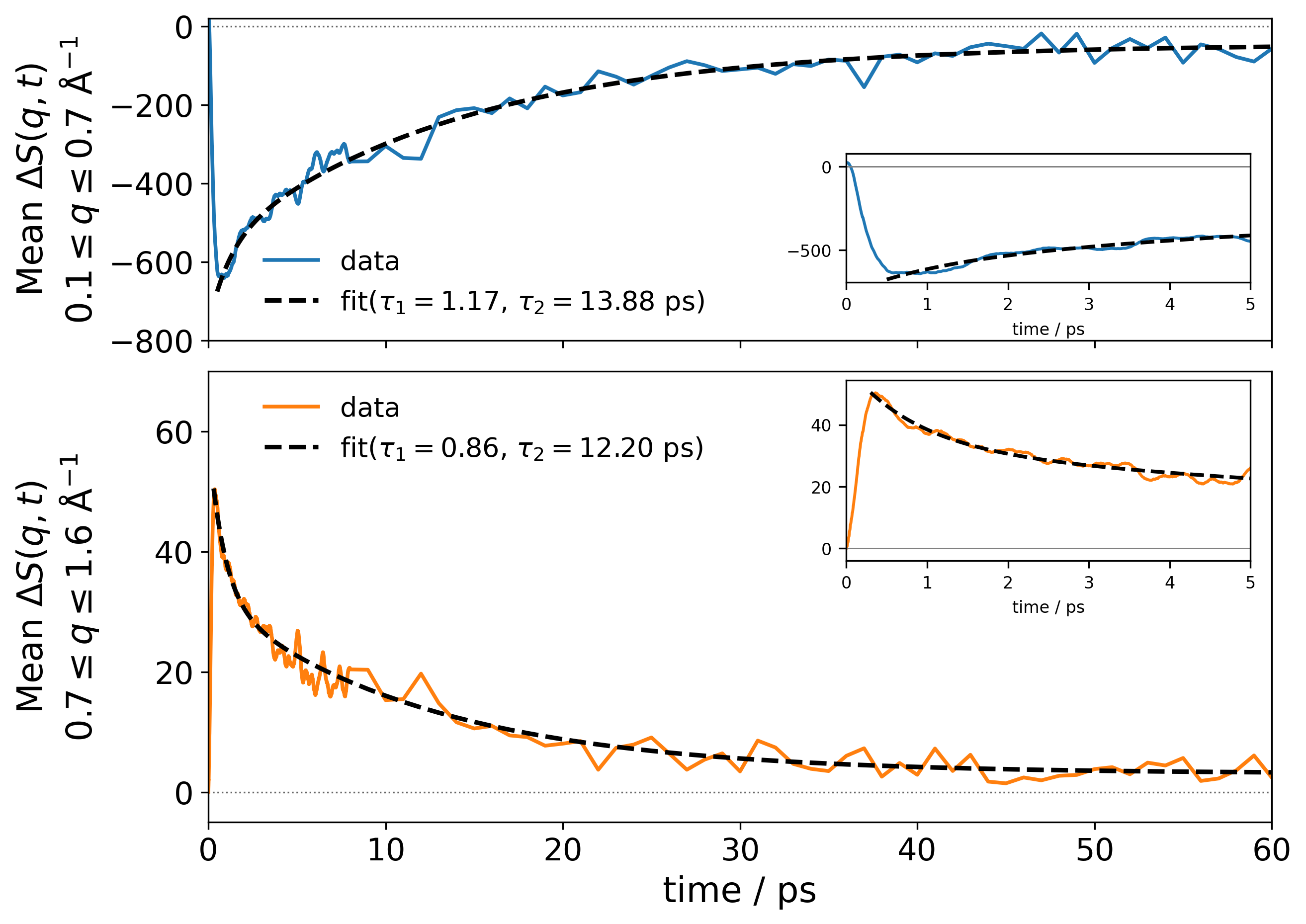}
    \caption{
    Time evolution of the solute--solvent difference X-ray scattering signal in Fig.~\ref{fig:heating_scattering} over selected \(q\) ranges, following vibrational excitation.
    Top: \(\Delta S_{\mathrm{uv}}(q,t)\) averaged over \(0.1\leq q\leq0.7~\text{\AA}^{-1}\).
    Bottom: The same contribution averaged over \(0.7\leq q\leq1.6~\text{\AA}^{-1}\).
    The insets show the first \(5~\mathrm{ps}\) in more detail.
    }
    \label{fig:heating_scattering_kinetics}
\end{figure}

We now analyze the nonequilibrium simulations of vibrational excitation of HTI and the corresponding calculated TR-XSS signal.

Figure~\ref{fig:heating_cavity_energy} shows the evolution of the solute and solvent kinetic energy.
Immediately after excitation, the solute kinetic energy undergoes a sharp drop, consistent with rapid conversion of part of the excess kinetic energy into intramolecular potential energy as the vibrational coordinates move away from their equilibrium values.
Subsequently, the solute kinetic energy decreases more gradually as energy is transferred to the solvent, and the kinetic energy of the latter increases.

The excluded volume initially increases from its equilibrium ground-state value of \(100.1~\text{\AA}^{3}\) to approximately \(106.5~\text{\AA}^{3}\), reaching a maximum at \(\sim0.4~\mathrm{ps}\) (Fig.~\ref{fig:heating_cavity_energy}, left).
This expansion is consistent with the hot solute having larger-amplitude vibrations, and therefore displacing the surrounding solvent molecules more.
As the solute cools by transferring energy to the solvent, the excluded volume contracts, with a biexponential fit yielding time constants of \(0.45\) and \(8.68~\mathrm{ps}\).
The excluded volume develops more slowly than the kinetic energy. The surrounding solvent requires time to move and rearrange in response to the increased vibrational amplitudes, explaining why the excluded volume reaches a maximum at only $\sim 0.4$ ps.
The contraction of the excluded volume is also expected to lag behind and proceed more slowly than the decrease in solute kinetic energy. The biexponential fit may not resolve a longer relaxation component. Beyond \(8~\mathrm{ps}\), configurations are sampled only every \(1~\mathrm{ps}\), and fluctuations in the volume limit the determination of slow dynamics.
The excluded volume remains slightly above its initial value at \(60~\mathrm{ps}\), suggesting that slower rearrangements may persist, although statistical uncertainty and a slightly increased thermal energy of the system could also contribute to this residual difference.

Figure~\ref{fig:heating_scattering} shows the solute--solvent and solvent--solvent difference X-ray scattering signals relative to the equilibrium ground state.
As in the charge-transfer simulations, the strongest amplitudes occur at low \(q\), where the two contributions exhibit similar features with opposite signs.
Figure~\ref{fig:heating_scattering_kinetics} shows the solute--solvent signal averaged over selected \(q\) intervals.
In the low-\(q\) interval, \(0.1\leq q\leq0.7~\text{\AA}^{-1}\), the negative signal reaches its largest absolute amplitude at approximately \(0.4\)--\(0.6~\mathrm{ps}\), close to the position of the maximum in the excluded volume, and subsequently recovers through a biexponential relaxation with time constants of \(1.17\) and \(13.88~\mathrm{ps}\). The higher-\(q\) interval, \(0.7\leq q\leq1.6~\text{\AA}^{-1}\), shows a positive signal with a similarly rapid onset followed by biexponential decay, with time constants of \(0.86\) and \(12.20~\mathrm{ps}\).
These values are somewhat closer to those obtained from the excluded volume than the low-\(q\) time constants, consistent with greater sensitivity to local restructuring of the first solvation shell.
The low-\(q\) signal instead may also reflect redistribution across the surrounding solvent layers, as for the charge-transfer simulations.

The positive low-\(q\) solvent--solvent contribution obtained in the heating simulations reproduces the experimental feature observed for HTI in acetonitrile by Mitterer \textit{et al.}\cite{Mitterer_2026}
The simulations thus support the assignment of this feature to changes in solvent packing caused by the increase in excluded volume of the vibrationally hot solute and the resulting displacement of surrounding solvent molecules.
At higher \(q\), the solvent--solvent contribution displays the bulk-solvent heating signature discussed above, with positive features around \(1.4\) and \(2.4~\text{\AA}^{-1}\) and a negative band near \(1.8~\text{\AA}^{-1}\), consistent with energy transfer into the acetonitrile bulk.\cite{Mitterer_2026,Kjaer_2013}

\section{Discussion}

The simulations performed here highlight both the potential of TR-XSS for elucidating photoinduced dynamics of organic molecules in solution and the challenges associated with interpreting the measured signals.
In studies of systems containing metal or halogen atoms, solute and solvent-cage dynamics are often interpreted using models based on a small number of selected nuclear coordinates.\cite{Markmann_2024,PowersRiggs_2024,Nimmrich_2023,Katayama_2023,Haldrup_2019}
The first TR-XSS studies of organic photoswitches highlight the importance of considering changes in solute and solvent nuclear distributions and structural rearrangements extending beyond a few selected internuclear coordinates.\cite{Kim_2026,Mitterer_2026}

HTI provides a convenient system for investigating polar solvation dynamics because its photoexcitation produces a large change in the molecular dipole moment.
The CAM-B3LYP calculations give an increase from approximately \(5.3\) to \(27.3~\mathrm{D}\) at the Franck--Condon geometry.
This strong electrostatic perturbation induces a large solvent rearrangement with an experimentally detectable scattering signature.\cite{Mitterer_2026}
The present simulations extend that interpretation of the previous TR-XSS measurements\cite{Mitterer_2026} by identifying some of the molecular motions and time scales contributing to the evolving signal, which could not be resolved experimentally.

\begin{figure*}[t]
    \centering
    \includegraphics[width=0.85\textwidth,height=0.38\textheight,keepaspectratio]{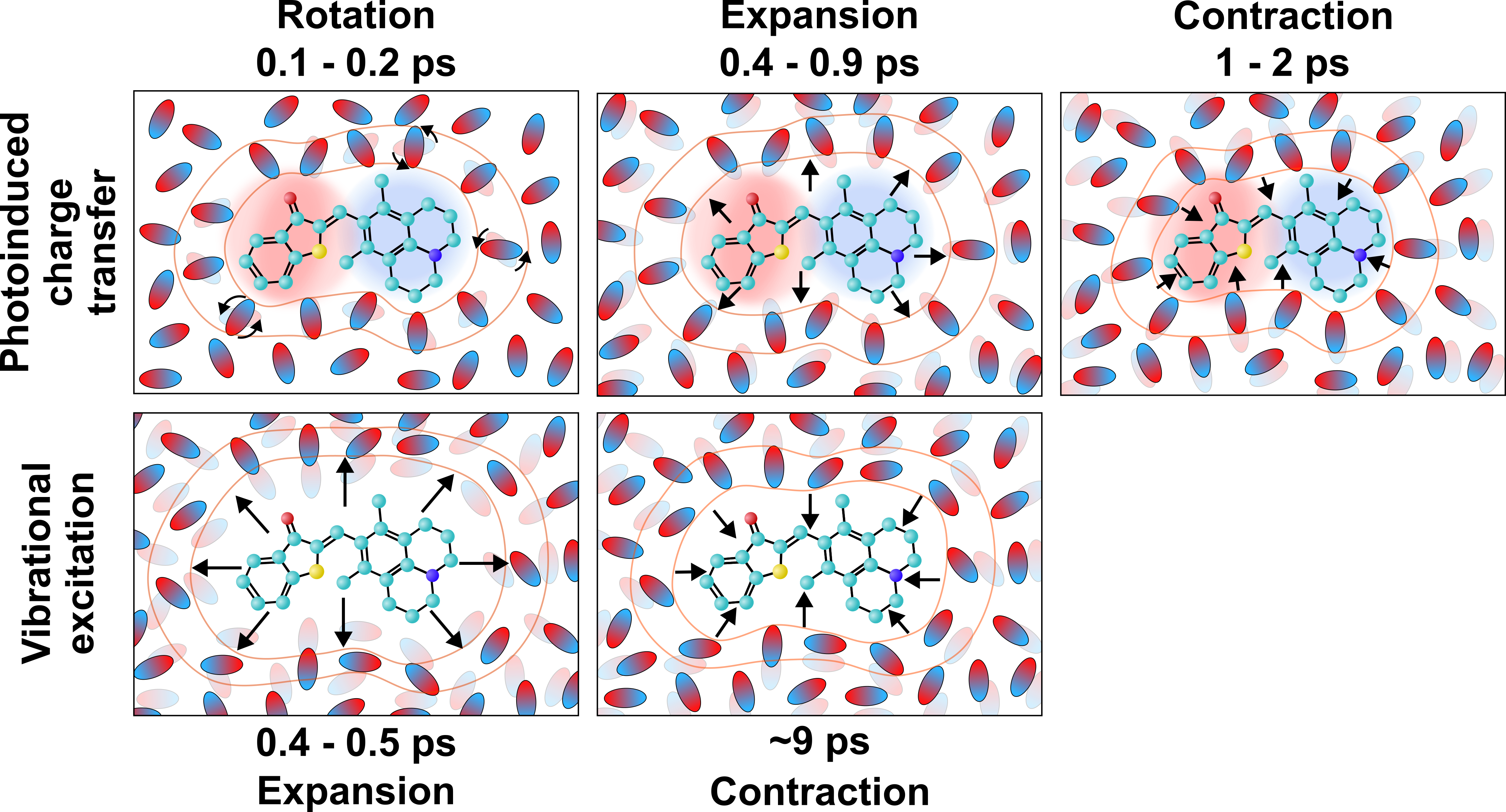}
    \caption{
    Schematic illustration of the time-dependent solvent response to photoinduced charge transfer (top) and instantaneous vibrational excitation of HTI (bottom) in acetonitrile, as characterized here using nonequilibrium molecular dynamics simulations.
    The sudden change in the solute charge distribution induces rapid reorientation of nearby acetonitrile molecules, followed by a transient expansion and slower contraction of the excluded volume.
    Vibrational excitation initially expands the excluded volume, which then contracts as the solute cools.
    Within HTI, blue denotes depletion of electron density in the julolidine donor and red denotes increased electron density in the thioindigo acceptor in the photoinduced charge transfer.
    For the acetonitrile molecules, red indicates the nitrogen end and blue the methyl end.
    All processes shown are reflected in the calculated time-dependent X-ray scattering signals in Figs.~\ref{fig:scattering_components} and~\ref{fig:heating_scattering}.
    }
    \label{fig:schema}
\end{figure*}

Figure~\ref{fig:schema} schematically summarizes the processes characterized in the nonequilibrium MD simulations following both photoinduced charge transfer and instantaneous vibrational excitation of the solute. The processes are reflected in the calculated time-dependent X-ray scattering signals in Figs.~\ref{fig:scattering_components} and~\ref{fig:heating_scattering}.

The solvent response to charge transfer is found to involve rapid reorientation of the solvent in the first solvation shell, with characteristic times of \(0.15\)--\(0.2~\mathrm{ps}\), as well as slower redistribution within and between solvation shells on picosecond time scales, with both translational and orientational contributions.
The excluded volume provides a convenient measure of the local solvent response.
Here, it is estimated as the cavity bounded by overlapping spheres centered on solvent atoms, and therefore reports primarily on the geometry and packing of the first solvation shell.
The time evolution of the excluded volume is indeed found to reflect the combined geometric response around the chromophore, involving an early contraction followed by a rebound, before a slower contraction occurs.

The solvation structural changes contribute differently across the measured \(q\) range.
Below approximately \(0.7~\text{\AA}^{-1}\), the signal reflects changes in local solvent packing as well as redistribution across surrounding shells, i.e. a more spatially extended rearrangement. The latter likely involve a large number of solvent molecules and diffusive reorganization.
At higher \(q\), the solute--solvent features around \(1~\text{\AA}^{-1}\) more closely follow the rapid reorientation, rebound, and subsequent relaxation of the first shell, while the solvent--solvent contribution exhibits the characteristic pattern of heating of the bulk solvent. Thus, the higher-$q$ signal reports mainly on local changes in atom-pair distances. Measuring both $q$-regions can therefore potentially distinguish local first-shell dynamics from redistribution across shells through comparison with simulated scattering signals.

A central difficulty is that the structural responses to charge transfer and vibrational excitation develop on overlapping time scales.
For HTI in acetonitrile, transient absorption measurements give an internal conversion time of \(2.6\pm0.5~\mathrm{ps}\).\cite{Mitterer_2026}
Internal conversion leads to solute heating and changes the solute charge distribution back toward that of the ground state.
Consequently, while the solvation shell is still relaxing in the excited state, the solvent molecules surrounding solute molecules that return to the ground state will experience both a new electrostatic perturbation and energy transfer from the hot solute.
The heating simulations show that vibrational excitation itself produces a significant cavity expansion, followed by contraction during cooling, contributing to the structural scattering signal.
The nonequilibrium simulations performed here characterize the solvation shell rearrangement to charge transfer and the structural changes associated with heating and cooling of the solute separately. However, they do not take into account that these processes might be coupled during the dynamics, a limitation of the approach employed here.

The earliest dynamics also warrant examination of the force-field approximation used here.
The initial nonexponential contraction and rebound within approximately \(0.2\)--\(0.4~\mathrm{ps}\) are consistent with rapid, torque-driven reorientation followed by a transient readjustment.
Their amplitude and timing may nevertheless depend on the description based on fixed charges employed here, which does not allow the electronic charge distributions of the solute and solvent to adapt dynamically to their changing environment. Polarizability might indeed ``soften'' the initial solvent response. Polarizable models provide a route to including this feedback.\cite{Nottoli_2021,Jing_2019}
Comparisons with polarizable models are therefore needed to establish to which the observed early-time dynamics might be a consequence of the lack of polarizability.

\section{Conclusions}

Nonequilibrium molecular dynamics simulations have been used to investigate separately the solvation dynamics upon photoinduced charge transfer and the response to vibrational excitation and cooling of the organic HTI chromophore in acetonitrile.
The underlying rearrangements have been characterized using time-resolved spatial distribution functions as well as the changes in the volume around the solute excluded to the solvent. The structural rearrangements have then been connected to calculated solute--solvent and solvent--solvent X-ray scattering signals.

Photoinduced charge transfer produces rapid first-shell reorientation with characteristic times of \(0.15\)--\(0.2~\mathrm{ps}\), accompanied by contraction of the excluded volume and a transient rebound within the first \(0.4~\mathrm{ps}\).
Slower solvent redistribution and further contraction follow on time scales of approximately \(1\)--\(2~\mathrm{ps}\).
Vibrational excitation instead expands the cavity, reaching a maximum near \(0.4~\mathrm{ps}\), followed by contraction as the solute cools over several to tens of picoseconds.
The calculated signals associate low-\(q\) features with redistribution across solvation shells as well as changes in the first shell, while higher-\(q\) features reflect local restructuring of the first shell more closely.
Higher-\(q\) features in the solvent-solvent scattering signal are associated to the the bulk heating response.
The present simulations reveal the wealth of information on local and large-amplitude solvent dynamics potentially accessible through TR-XSS measurements of organic molecules photoexcited in solution, and show that interpreting these signals requires accounting for changes in solvent packing induced by vibrational heating and cooling of the solute.

Further progress requires efficient models that simultaneously account for changes in populations of electronic states, energy transfer, and solvent relaxation\cite{Ventura_2022} while retaining the extensive sampling needed for converged scattering calculations.
We plan to incorporate solute and solvent polarizability and investigate the molecular pathways of energy transfer.
Structural distributions alone cannot always establish how much energy enters solvent rotations or librations, translations, and intramolecular vibrations, as is the case for the vibrational cooling of the solute.
An analyses of the energy fluxes based on power and work offer a way to resolve the different pathways and connect the observed restructuring to the mechanisms of energy transfer,\cite{Birgisson_2025,Rey_2015} and this strategy will be considered in future work.

\begin{acknowledgments}
E.S. and G.L. acknowledge support from the Icelandic Research Fund (Grant Nos. 239678, 2511544).
S.P. and G.L. acknowledge support from the ERC under the European Union's Horizon Europe research and innovation programme (Grant No. 101166044, project NEXUS).
Views and opinions expressed are however those of the author(s) only and do not necessarily reflect those of the European Union or ERC Executive Agency.
Neither the European Union nor the granting authority can be held responsible for them. A.O.D. and K.B.M. gratefully acknowledge support from the Independent Research Fund Denmark (Grant No. 1127-00382B).
\end{acknowledgments}

\section*{Author Declarations}
The authors have no conflicts to disclose.

\section*{DATA AVAILABILITY STATEMENT}
Data supporting the findings of this study are available from the corresponding author upon reasonable request.

\clearpage


\section{References}
\bibliography{aipsamp}

@PREAMBLE{
 "\providecommand{\noopsort}[1]{}" 
 # "\providecommand{\singleletter}[1]{#1}%" 
}

@misc{grsq,
    author = {Dohn, Asmus Ougaard},
    title = {\texttt{grsq}: Structure factor and scattering from radial distribution functions},
    howpublished = {\url{https://pypi.org/project/grsq/}},
    note = {Accessed October 5, 2026}
}

@article{Bondi_1964,
   author = {A Bondi},
   doi = {10.1021/j100785a001},
   issn = {0022-3654},
   journal = {J. Phys. Chem.},
   month = {3},
   pages = {441--451},
   title = {van der Waals Volumes and Radii},
   volume = {68},
   url = {https://doi.org/10.1021/j100785a001},
   year = {1964}
}

@article{Debye_1915,
    author  = {Debye, P.},
    title   = {Zerstreuung von R{\"o}ntgenstrahlen},
    journal = {Ann. Phys.},
    volume  = {351},
    number  = {6},
    pages   = {809--823},
    year    = {1915},
    doi     = {10.1002/andp.19153510606}
}

@article{Gaff_2004,
   author = {Junmei Wang and Romain M Wolf and James W Caldwell and Peter A Kollman and David A Case},
   doi = {10.1002/jcc.20035},
   journal = {J. Comput. Chem.},
   pages = {1157--1174},
   title = {Development and Testing of a General Amber Force Field},
   volume = {25},
   url = {https://onlinelibrary.wiley.com/doi/10.1002/jcc.20035},
   year = {2004}
}

@article{Openmm_2024,
   author = {Peter Eastman and Raimondas Galvelis and Raúl P. Peláez and Charlles R.A. Abreu and Stephen E. Farr and Emilio Gallicchio and Anton Gorenko and Michael M. Henry and Frank Hu and Jing Huang and Andreas Krämer and Julien Michel and Joshua A. Mitchell and Vijay S. Pande and João PGLM Rodrigues and Jaime Rodriguez-Guerra and Andrew C. Simmonett and Sukrit Singh and Jason Swails and Philip Turner and Yuanqing Wang and Ivy Zhang and John D. Chodera and Gianni De Fabritiis and Thomas E. Markland},
   doi = {10.1021/acs.jpcb.3c06662},
   issn = {15205207},
   number = {1},
   journal = {J. Phys. Chem. B},
   month = {1},
   pages = {109--116},
   pmid = {38154096},
   publisher = {American Chemical Society},
   title = {OpenMM 8: Molecular Dynamics Simulation with Machine Learning Potentials},
   volume = {128},
   year = {2024}
}

@article{Fleming_1989,
author = {Mark Maroncelli  and Jean MacInnis  and Graham R. Fleming},
title = {Polar Solvent Dynamics and Electron-Transfer Reactions},
journal = {Science},
volume = {243},
number = {4899},
pages = {1674--1681},
year = {1989},
doi = {10.1126/science.243.4899.1674},
URL = {https://www.science.org/doi/abs/10.1126/science.243.4899.1674},
}

@article{Bagchi_2010,
   author = {Biman Bagchi and Biman Jana},
   doi = {10.1039/b902048a},
   issn = {14604744},
   issue = {6},
   journal = {Chem. Soc. Rev.},
   month = {5},
   pages = {1936--1954},
   title = {Solvation dynamics in dipolar liquids},
   volume = {39},
   year = {2010}
}

@article{Choi_2022,
   author = {Eun Hyuk Choi and Yunbeom Lee and Jun Heo and Hyotcherl Ihee},
   doi = {10.1039/d2sc00502f},
   issn = {20416539},
   issue = {29},
   journal = {Chem. Sci.},
   month = {6},
   pages = {8457--8490},
   publisher = {Royal Society of Chemistry},
   title = {Reaction dynamics studied via femtosecond X-ray liquidography at X-ray free-electron lasers},
   volume = {13},
   year = {2022}
}

@article{Biasin_2021,
   author = {Elisa Biasin and Zachary W. Fox and Amity Andersen and Kathryn Ledbetter and Kasper S. Kjær and Roberto Alonso-Mori and Julia M. Carlstad and Matthieu Chollet and James D. Gaynor and James M. Glownia and Kiryong Hong and Thomas Kroll and Jae Hyuk Lee and Chelsea Liekhus-Schmaltz and Marco Reinhard and Dimosthenis Sokaras and Yu Zhang and Gilles Doumy and Anne Marie March and Stephen H. Southworth and Shaul Mukamel and Kelly J. Gaffney and Robert W. Schoenlein and Niranjan Govind and Amy A. Cordones and Munira Khalil},
   doi = {10.1038/s41557-020-00629-3},
   issn = {17554349},
   issue = {4},
   journal = {Nat. Chem.},
   month = {4},
   pages = {343-349},
   pmid = {33589787},
   publisher = {Nature Research},
   title = {Direct observation of coherent femtosecond solvent reorganization coupled to intramolecular electron transfer},
   volume = {13},
   year = {2021}
}

@article{Horng_1995,
   author = {M. L. Horng and J. A. Gardecki and A. Papazyan and M. Maroncelli},
   doi = {10.1021/j100048a004},
   journal = {J. Phys. Chem.},
   pages = {17311--17337},
   title = {Subpicosecond Measurements of Polar Solvation Dynamics: Coumarin 153 Revisited},
   volume = {99},
   year = {1995}
}

@article{Ki_2021,
    author = {Ki, Hosung and Choi, Seungjoo and Kim, Jungmin and Choi, Eun Hyuk and Lee, Seonggon and Lee, Yunbeom and Yoon, Kihwan and Ahn, Chi Woo and Ahn, Doo-Sik and Lee, Jae Hyuk and Park, Jaeku and Eom, Intae and Kim, Minseok and Chun, Sae Hwan and Kim, Joonghan and Ihee, Hyotcherl and Kim, Jeongho},
    title = {Optical
Kerr Effect of Liquid Acetonitrile Probed
by Femtosecond Time-Resolved X‑ray Liquidography},
    journal = {J. Am. Chem. Soc.},
    volume = {143},
    number = {35},
    pages = {14261--14273},
    year = {2021},
    month = {09},
    issn = {0002-7863},
    doi = {10.1021/jacs.1c06088},
}

@article{Mitterer_2026,
   author = {Kerstin M Mitterer and Elli Selenius and Morten L Haubro and Magnus A H Christiansen and Verena Markmann and Bianca L Hansen and Mikkel Krell-Jørgensen and Joseph G F Hoock and Victor Lorentzen and Emma V Beale and Philip J M Johnson and David J Gosztola and Claudio Cirelli and Camila Bacellar and Asmus O Dohn and Luca Laraia and Klaus B Møller and Kristoffer Haldrup and Gianluca Levi and Martin M Nielsen},
   journal = {Nat. Commun.},
   volume={17},
   year = {2026},
   pages = {5080},
   doi = {10.1038/s41467-026-71635-1},
   title = {Tracking polar solvation dynamics of a photoexcited organic chromophore with ultrafast X-ray scattering},
   url = {https://doi.org/10.1038/s41467-026-71635-1}
}

@article{Thompson_2024,
   author = {Niklas B. Thompson and Karen L. Mulfort and David M. Tiede},
   doi = {10.1107/S2052252524003282},
   issn = {20522525},
   issue = {Pt 3},
   journal = {IUCrJ},
   month = {5},
   pages = {423--433},
   publisher = {International Union of Crystallography},
   title = {Toward a quantitative description of solvation structure: a framework for differential solution scattering measurements},
   volume = {11},
   year = {2024}
}

@article{Rey_2020,
   author = {Rossend Rey and James T. Hynes},
   doi = {10.1021/acs.jpcb.0c05706},
   issn = {15205207},
   issue = {35},
   journal = {J. Phys. Chem. B},
   month = {9},
   pages = {7668--7681},
   pmid = {32790403},
   publisher = {American Chemical Society},
   title = {Solvation Dynamics in Water. 4. On the Initial Regime of Solvation Relaxation},
   volume = {124},
   year = {2020}
}

@article{Papai_2019,
  author = {P{\'a}pai, M{\'a}ty{\'a}s and Abedi, Mostafa and Levi, Gianluca and Biasin, Elisa and Nielsen, Martin M. and M{\o}ller, Klaus B.},
  title = {Theoretical Evidence of Solvent-Mediated Excited-State Dynamics in a Functionalized Iron Sensitizer},
  journal = {J. Phys. Chem. C},
  year = {2019},
  volume = {123},
  number = {4},
  pages = {2056--2065},
  doi = {10.1021/acs.jpcc.8b10768}
}

@article{Ingrosso_2005,
   author = {Ingrosso, Francesca and Ladanyi, Branka M. and Mennucci, Benedetta and Elola, M. Dolores and Tomasi, Jacopo},
   journal = {J. Phys. Chem. B},
   year = {2005},
   volume={109},
   number={8},
   pages={3553--3564},
   doi = {10.1021/jp0456032},
   title = {Solvation Dynamics in Acetonitrile: A Study Incorporating Solute Electronic Response and Nuclear Relaxation},
}

@article{Dohn_2023,
   author = {Asmus O Dohn and Verena Markmann and A. Nimmrich and Kristoffer Haldrup and Klaus B Møller and Martin M Nielsen},
   journal = {J. Chem. Phys.},
   year = {2023},
   volume={159},
   number={12},
   pages={124115},
   doi = {10.1063/5.0164365},
   title = {Eliminating ﬁnite-size effects on the calculation of
x-ray scattering from molecular dynamics simulations},
}

@article{Lorch_1969,
doi = {10.1088/0022-3719/2/2/305},
url = {https://doi.org/10.1088/0022-3719/2/2/305},
year = {1969},
month = {feb},
publisher = {},
volume = {2},
number = {2},
pages = {229},
author = {E Lorch},
title = {Neutron diffraction by germania, silica and radiation-damaged silica glasses},
journal = {J. Phys. C: Solid State Phys.},
}

@article{Oang_2023,
    author = {Oang, Key Young and Park, Sungjun and Moon, Jiwon and Park, Eunji and Lee, Hyun Kyung and Sato, Tokushi and Nozawa, Shunsuke and Adachi, Shin-ichi and Kim, Joonghan and Kim, Jeongho and Sohn, Jeong-Hun and Ihee, Hyotcherl},
    title = {Extracting Kinetics and Thermodynamics of Molecules
without Heavy Atoms via Time-Resolved Solvent Scattering Signals},
    journal = {J. Phys. Chem. Lett.},
    volume = {14},
    number = {13},
    pages = {3103--3110},
    year = {2023},
    month = {04},
    issn = {1948-7185},
}

@article{Dohn_2015,
  title={On the calculation of x-ray scattering signals from pairwise radial distribution functions},
  author={Dohn, Asmus O and Biasin, Elisa and Haldrup, Kristoffer and Nielsen, Martin M and Henriksen, Niels E and M{\o}ller, Klaus B},
  journal={J. Phys. B: At. Mol. Opt. Phys.},
  volume={48},
  number={24},
  pages={244010},
  year={2015},
  publisher={IOP Publishing},
  doi={10.1088/0953-4075/48/24/244010},
}

@book{Warren_1990,
  author    = {Warren, B. E.},
  title     = {X-Ray Diffraction},
  publisher = {Dover Publications},
  address   = {Mineola, NY},
  year      = {1990}
}

@article{Dohn_2016b,
  title={Corrigendum: On the calculation of x-ray scattering signals from pairwise radial distribution functions (2015 J. Phys. B: At. Mol. Opt. Phys. 48 244010)},
  author={Dohn, Asmus Ougaard and Biasin, Elisa and Haldrup, Kristoffer and Nielsen, Martin Meedom and Henriksen, Niels Engholm and M{\o}ller, Klaus Braagaard},
  journal={J. Phys. B: At. Mol. Opt. Phys.},
  volume={49},
  number={5},
  pages={059501},
  year={2016},
  publisher={IOP Publishing},
  doi={10.1088/0953-4075/49/5/059501}
}

@article{Ganguly_2013,
    author = {Ganguly, Pritam and van der Vegt, Nico F. A.},
    title = {Convergence of Sampling Kirkwood–Buff Integrals
of Aqueous Solutions with Molecular Dynamics Simulations},
    journal = {J. Chem. Theory Comput.},
    volume = {9},
    number = {3},
    pages = {1347--1355},
    year = {2013},
    month = {03},
    issn = {1549-9618},
    doi = {10.1021/ct301017q},
    url = {https://doi.org/10.1021/ct301017q},
}

@article{Rondi_2015,
  author = {Rondi, Ariana and Rodriguez, Yuseff and Feurer, Thomas and Cannizzo, Andrea},
  title = {Solvation-Driven Charge Transfer and Localization in Metal Complexes},
  journal = {Acc. Chem. Res.},
  year = {2015},
  volume = {48},
  number = {5},
  pages = {1432--1440},
  doi = {10.1021/ar5003939}
}

@article{Kumpulainen_2017_review,
  author = {Kumpulainen, Tatu and Lang, Bernhard and Rosspeintner, Arnulf and Vauthey, Eric},
  title = {Ultrafast Elementary Photochemical Processes of Organic Molecules in Liquid Solution},
  journal = {Chem. Rev.},
  year = {2017},
  volume = {117},
  number = {16},
  pages = {10826--10939},
  doi = {10.1021/acs.chemrev.6b00491}
}

@article{Dereka_2020,
  author = {Dereka, Bogdan and Svechkarev, Denis and Rosspeintner, Arnulf and Aster, Alexander and Lunzer, Markus and Liska, Robert and Mohs, Aaron M. and Vauthey, Eric},
  title = {Solvent tuning of photochemistry upon excited-state symmetry breaking},
  journal = {Nat. Commun.},
  year = {2020},
  volume = {11},
  pages = {1925},
  doi = {10.1038/s41467-020-15681-3}
}

@article{Rosenthal_1991,
  author = {Rosenthal, Sandra J. and Xie, Xiaoliang and Du, Mei and Fleming, Graham R.},
  title = {Femtosecond solvation dynamics in acetonitrile: Observation of the inertial contribution to the solvent response},
  journal = {J. Chem. Phys.},
  year = {1991},
  volume = {95},
  number = {6},
  pages = {4715--4718},
  doi = {10.1063/1.461742}
}

@article{Jimenez_1994,
  author = {Jimenez, Ralph and Fleming, Graham R. and Kumar, P. V. and Maroncelli, Mark},
  title = {Femtosecond solvation dynamics of water},
  journal = {Nature},
  year = {1994},
  volume = {369},
  pages = {471--473},
  doi = {10.1038/369471a0}
}

@article{Lee_2024,
   author = {Lee, Yunbeom and Oang, Key Young and Kim, Doyeong and Ihee, Hyotcherl},
   title = {A comparative review of time-resolved {x-ray} and electron scattering to probe structural dynamics},
   journal = {Struct. Dyn.},
   year = {2024},
   volume = {11},
   number = {3},
   pages = {031301},
   doi = {10.1063/4.0000249}
}

@article{Lee_2013,
   author = {Lee, Jae Hyuk and Wulff, Michael and Bratos, Savo and Petersen, Jakob and Guerin, Laurent and Leicknam, Jean-Claude and Cammarata, Marco and Kong, Qingyu and Kim, Jeongho and Møller, Klaus B. and Ihee, Hyotcherl},
   title = {Filming the Birth of Molecules and Accompanying Solvent Rearrangement},
   journal = {J. Am. Chem. Soc.},
   year = {2013},
   volume = {135},
   number = {8},
   pages = {3255--3261},
   doi = {10.1021/ja312513w}
}

@article{vanDriel_2016,
  author = {van Driel, Tim B. and Kj{\ae}r, Kasper S. and Hartsock, Robert W. and Dohn, Asmus O. and Harlang, Tobias and Chollet, Matthieu and Christensen, Morten and Gawelda, Wojciech and Henriksen, Niels E. and Kim, Jong Goo and Haldrup, Kristoffer and Kim, Kyung Hwan and Ihee, Hyotcherl and Kim, Jeongho and Lemke, Henrik and Sun, Zheng and Sundstr{\"o}m, Villy and Zhang, Wenkai and Zhu, Diling and M{\o}ller, Klaus B. and Nielsen, Martin M. and Gaffney, Kelly J.},
  title = {Atomistic characterization of the active-site solvation dynamics of a model photocatalyst},
  journal = {Nat. Commun.},
  year = {2016},
  volume = {7},
  pages = {13678},
  doi = {10.1038/ncomms13678}
}

@article{Haldrup_2016,
   author = {Haldrup, Kristoffer and Gawelda, Wojciech and Abela, Rafael and Alonso-Mori, Roberto and Bergmann, Uwe and Bordage, Amélie and Cammarata, Marco and Canton, Sophie E. and Dohn, Asmus O. and van Driel, Tim Brandt and Fritz, David M. and Galler, Andreas and Glatzel, Pieter and Harlang, Tobias and Kjær, Kasper S. and Lemke, Henrik T. and Møller, Klaus B. and Németh, Zoltán and Pápai, Mátyás and Sas, Norbert and Uhlig, Jens and Zhu, Diling and Vankó, György and Sundström, Villy and Nielsen, Martin M. and Bressler, Christian},
   title = {Observing Solvation Dynamics with Simultaneous Femtosecond {X-ray} Emission Spectroscopy and {X-ray} Scattering},
   journal = {J. Phys. Chem. B},
   year = {2016},
   volume = {120},
   number = {6},
   pages = {1158--1168},
   doi = {10.1021/acs.jpcb.5b12471}
}

@article{Nimmrich_2023,
   author = {Nimmrich, Amke and Panman, Matthijs R. and Berntsson, Oskar and Biasin, Elisa and Niebling, Stephan and Petersson, Jonas and Hoernke, Maria and Björling, Alexander and Gustavsson, Emil and van Driel, Tim B. and Dohn, Asmus O. and Laursen, Mads and Zederkof, Diana B. and Tono, Kensuke and Katayama, Tetsuo and Owada, Shigeki and Nielsen, Martin M. and Davidsson, Jan and Uhlig, Jens and Hub, Jochen S. and Haldrup, Kristoffer and Westenhoff, Sebastian},
   title = {Solvent-Dependent Structural Dynamics in the Ultrafast Photodissociation Reaction of Triiodide Observed with Time-Resolved {X-ray} Solution Scattering},
   journal = {J. Am. Chem. Soc.},
   year = {2023},
   volume = {145},
   number = {29},
   pages = {15754--15765},
}

@article{Heo_2022,
   author = {Heo, Jun and Kim, Jong Goo and Choi, Eun Hyuk and Ki, Hosung and Ahn, Doo-Sik and Kim, Jungmin and Lee, Seonggon and Ihee, Hyotcherl},
   title = {Determining the charge distribution and the direction of bond cleavage with femtosecond anisotropic {x-ray} liquidography},
   journal = {Nat. Commun.},
   year = {2022},
   volume = {13},
   number = {1},
   pages = {522},
   doi = {10.1038/s41467-022-28168-0}
}

@article{Markmann_2024,
   author = {Markmann, Verena and Pan, Jaysree and Hansen, Bianca L. and Haubro, Morten L. and Nimmrich, Amke and Lenzen, Philipp and Levantino, Matteo and Katayama, Tetsuo and Adachi, Shin-ichi and Gorski-Bilke, Simone and Temps, Friedrich and Dohn, Asmus O. and Møller, Klaus B. and Nielsen, Martin M. and Haldrup, Kristoffer},
   title = {Real-time structural dynamics of the ultrafast solvation process around photo-excited aqueous halides},
   journal = {Chem. Sci.},
   year = {2024},
   volume = {15},
   number = {29},
   pages = {11391--11401},
   doi = {10.1039/d4sc01912a}
}

@article{Vester_2022,
   author = {Vester, Peter and Kubicek, Katharina and Alonso-Mori, Roberto and Assefa, Tadesse and Biasin, Elisa and Christensen, Morten and Dohn, Asmus O. and van Driel, Tim B. and Galler, Andreas and Gawelda, Wojciech and Harlang, Tobias C. B. and Henriksen, Niels E. and Kjær, Kasper S. and Kuhlman, Thomas S. and Németh, Zoltán and Nurekeyev, Zhangatay and Pápai, Mátyás and Rittman, Jochen and Vankó, György and Yavas, Hasan and Zederkof, Diana B. and Bergmann, Uwe and Nielsen, Martin M. and Møller, Klaus B. and Haldrup, Kristoffer and Bressler, Christian},
   title = {Tracking structural solvent reorganization and recombination dynamics following {e$^{-}$} photoabstraction from aqueous {I$^{-}$} with femtosecond {x-ray} spectroscopy and scattering},
   journal = {J. Chem. Phys.},
   year = {2022},
   volume = {157},
   number = {22},
   pages = {224201},
   doi = {10.1063/5.0107224}
}

@article{Kahraman_2026,
   author = {Kahraman, Abdullah and Sachs, Michael and Ghosh, Soumen and Poulter, Benjamin I. and Sucre-Rosales, Estefanía and Ryland, Elizabeth S. and Garratt, Douglas and Raj, Sumana L. and Powers-Riggs, Natalia and Kundu, Subhradip and Hampton, Christina Y. and Hoffman, David J. and Coslovich, Giacomo and Dakovski, Georgi L. and Kramer, Patrick L. and Chollet, Matthieu and Alonso-Mori, Roberto and van Driel, Tim B. and Lee, Sang-Jun and Kunnus, Kristjan and Cordones, Amy A. and Schoenlein, Robert W. and Vauthey, Eric and Andersen, Amity and Govind, Niranjan and Larsen, Christopher B. and Biasin, Elisa},
   title = {Electronic and solvent reorganization in proton-coupled electron transfer captured by ultrafast {X-rays}},
   journal = {Nat. Commun.},
   year = {2026},
   volume = {17},
   number = {1},
   pages = {9053},
   doi = {10.1038/s41467-026-75943-4}
}

@article{Kim_2026,
   author = {Kim, Jungmin and Ki, Hosung and Lee, Seonggon and Segalina, Alekos and Lee, Yunbeom and Ihee, Hyotcherl},
   title = {{X-ray} liquidography decodes complex motions in azobenzene isomerization},
   journal = {Nature},
   year = {2026},
}

@article{Rey_2015,
  author = {Rey, Rossend and Hynes, James T.},
  title = {Solvation Dynamics in Liquid Water. 1. Ultrafast Energy Fluxes},
  journal = {J. Phys. Chem. B},
  year = {2015},
  volume = {119},
  number = {24},
  pages = {7558--7570},
  doi = {10.1021/jp5113922}
}

@article{Yanai_2004,
  author = {Yanai, Takeshi and Tew, David P. and Handy, Nicholas C.},
  title = {A new hybrid exchange-correlation functional using the Coulomb-attenuating method ({CAM-B3LYP})},
  journal = {Chem. Phys. Lett.},
  year = {2004},
  volume = {393},
  number = {1-3},
  pages = {51--57},
  doi = {10.1016/j.cplett.2004.06.011}
}

@article{Zheng_2011,
  author = {Zheng, Jingjing and Xu, Xin and Truhlar, Donald G.},
  title = {Minimally augmented {Karlsruhe} basis sets},
  journal = {Theor. Chem. Acc.},
  year = {2011},
  volume = {128},
  pages = {295--305},
  doi = {10.1007/s00214-010-0846-z}
}

@article{Weigend_2005,
  author = {Weigend, Florian and Ahlrichs, Reinhart},
  title = {Balanced basis sets of split valence, triple zeta valence and quadruple zeta valence quality for {H} to {Rn}: Design and assessment of accuracy},
  journal = {Phys. Chem. Chem. Phys.},
  year = {2005},
  volume = {7},
  number = {18},
  pages = {3297--3305},
  doi = {10.1039/B508541A}
}

@article{Breneman_1990,
  author = {Breneman, Curt M. and Wiberg, Kenneth B.},
  title = {Determining atom-centered monopoles from molecular electrostatic potentials. The need for high sampling density in formamide conformational analysis},
  journal = {J. Comput. Chem.},
  year = {1990},
  volume = {11},
  number = {3},
  pages = {361--373},
  doi = {10.1002/jcc.540110311}
}

@article{Cossi_2003,
  author = {Cossi, Maurizio and Rega, Nadia and Scalmani, Giovanni and Barone, Vincenzo},
  title = {Energies, structures, and electronic properties of molecules in solution with the {C-PCM} solvation model},
  journal = {J. Comput. Chem.},
  year = {2003},
  volume = {24},
  number = {6},
  pages = {669--681},
  doi = {10.1002/jcc.10189}
}

@article{Barone_1998,
  author = {Barone, Vincenzo and Cossi, Maurizio},
  title = {Quantum calculation of molecular energies and energy gradients in solution by a conductor solvent model},
  journal = {J. Phys. Chem. A},
  year = {1998},
  volume = {102},
  number = {11},
  pages = {1995--2001},
  doi = {10.1021/jp9716997}
}

@misc{Restaino_2026,
  author = {Restaino, Lorenzo and Gamboni, Giulia and Selenius, Elli and Levi, Gianluca},
  title = {Orbital-optimized density functional calculations of excited electronic states: Recent advances and perspectives},
  howpublished = {arXiv:2606.13654},
  year = {2026},
  doi = {10.48550/arXiv.2606.13654},
  url = {https://arxiv.org/abs/2606.13654}
}

@article{Schmerwitz_2026,
  author = {Schmerwitz, Yorick L. A. and Selenius, Elli and Levi, Gianluca},
  title = {Freeze-and-Release Direct Optimization Method for Variational Calculations of Excited Electronic States},
  journal = {J. Chem. Theory Comput.},
  year = {2026},
  volume = {22},
  number = {7},
  pages = {3571--3584},
  doi = {10.1021/acs.jctc.5c01974}
}

@article{Levi_2020,
  author = {Levi, Gianluca and Ivanov, Aleksei V. and J{\'o}nsson, Hannes},
  title = {Variational Density Functional Calculations of Excited States via Direct Optimization},
  journal = {J. Chem. Theory Comput.},
  year = {2020},
  volume = {16},
  number = {11},
  pages = {6968--6982},
  doi = {10.1021/acs.jctc.0c00597}
}

@article{Neese2012,
  title={The ORCA program system},
  author={Neese, Frank},
  journal={Wiley Interdiscip. Rev. Comput. Mol. Sci.},
  volume={2},
  number={1},
  pages={73--78},
  year={2012},
  publisher={Wiley Online Library},
  doi={10.1002/wcms.81}
}

@article{Neese2022,
  title={Software update: The ORCA program system—Version 5.0},
  author={Neese, Frank},
  journal={Wiley Interdiscip. Rev. Comput. Mol. Sci.},
  volume={12},
  number={5},
  pages={e1606},
  year={2022},
  publisher={Wiley Online Library},
  doi={10.1002/wcms.1606}
}

@article{Kumar_1995,
  author = {Kumar, P. V. and Maroncelli, Mark},
  title = {Polar solvation dynamics of polyatomic solutes: Simulation studies in acetonitrile and methanol},
  journal = {J. Phys. Chem.},
  year = {1995},
  volume = {103},
  number = {8},
  pages = {3038--3060},
  doi = {10.1063/1.470493}
}

@article{Maroncelli_1991,
  author = {Maroncelli, Mark},
  title = {Computer simulations of solvation dynamics in acetonitrile},
  journal = {J. Chem. Phys.},
  year = {1991},
  volume = {94},
  number = {3},
  pages = {2084--2103},
  doi = {10.1063/1.459932}
}

@article{Kjaer_2013,
  author = {Kj{\ae}r, Kasper Skov and van Driel, Tim B. and Kehres, Jan and Haldrup, Kristoffer and Khakhulin, Dmitry and Bechgaard, Klaus and Cammarata, Marco and Wulff, Michael and S{\o}rensen, Thomas Just and Nielsen, Martin M.},
  title = {Introducing a standard method for experimental determination of the solvent response in laser pump, {X-ray} probe time-resolved wide-angle {X-ray} scattering experiments on systems in solution},
  journal = {Phys. Chem. Chem. Phys.},
  year = {2013},
  volume = {15},
  number = {36},
  pages = {15003--15016},
  doi = {10.1039/C3CP50751C}
}

@article{Haldrup_2019,
  author = {Haldrup, Kristoffer and Levi, Gianluca and Biasin, Elisa and Vester, Peter and Laursen, Mads Goldschmidt and Beyer, Frederik and Kj{\ae}r, Kasper Skov and van Driel, Tim Brandt and Harlang, Tobias and Dohn, Asmus O. and Hartsock, Robert J. and Nelson, Silke and Glownia, James M. and Lemke, Henrik T. and Christensen, Morten and Gaffney, Kelly J. and Henriksen, Niels E. and M{\o}ller, Klaus B. and Nielsen, Martin M.},
  title = {Ultrafast {X-Ray} Scattering Measurements of Coherent Structural Dynamics on the Ground-State Potential Energy Surface of a Diplatinum Molecule},
  journal = {Phys. Rev. Lett.},
  year = {2019},
  volume = {122},
  number = {6},
  pages = {063001},
  doi = {10.1103/PhysRevLett.122.063001}
}

@article{PowersRiggs_2024,
  author = {Powers-Riggs, Natalia E. and Birgisson, Benedikt O. and Raj, Sumana L. and Biasin, Elisa and Lenzen, Philipp and Zederkof, Diana Bregenholt and Haubro, Morten and Tveiten, Dagr{\'u}n K. V. and Hartsock, Robert W. and van Driel, Tim B. and Kunnus, Kristjan and Chollet, Matthieu and Robinson, Joseph S. and Nelson, Silke and Forbes, Ruaridh and Haldrup, Kristoffer and Pedersen, Kasper S. and Levi, Gianluca and Dohn, Asmus Ougaard and J{\'o}nsson, Hannes and M{\o}ller, Klaus B. and Natan, Adi and Nielsen, Martin Meedom and Gaffney, Kelly J.},
  title = {Characterization of Deformational Isomerization Potential and Interconversion Dynamics with Ultrafast {X-ray} Solution Scattering},
  journal = {J. Am. Chem. Soc.},
  year = {2024},
  volume = {146},
  number = {20},
  pages = {13962--13973},
  doi = {10.1021/jacs.4c00817}
}

@article{Katayama_2023,
  author = {Katayama, Tetsuo and Choi, Tae-Kyu and Khakhulin, Dmitry and Dohn, Asmus O. and Milne, Christopher J. and Vank{\'o}, Gy{\"o}rgy and N{\'e}meth, Zolt{\'a}n and Lima, Frederico A. and Szlachetko, Jakub and Sato, Tokushi and Nozawa, Shunsuke and Adachi, Shin-ichi and Yabashi, Makina and Penfold, Thomas J. and Gawelda, Wojciech and Levi, Gianluca},
  title = {Atomic-scale observation of solvent reorganization influencing photoinduced structural dynamics in a copper complex photosensitizer},
  journal = {Chem. Sci.},
  year = {2023},
  volume = {14},
  number = {10},
  pages = {2572--2584},
  doi = {10.1039/D2SC06600A}
}

@article{Nottoli_2021,
  author = {Nottoli, Michele and Cupellini, Lorenzo and Lipparini, Filippo and Granucci, Giovanni and Mennucci, Benedetta},
  title = {Multiscale Models for Light-Driven Processes},
  journal = {Annu. Rev. Phys. Chem.},
  year = {2021},
  volume = {72},
  pages = {489--513},
  doi = {10.1146/annurev-physchem-090419-104031}
}

@article{Jing_2019,
  author = {Jing, Zhifeng and Liu, Chengwen and Cheng, Sara Y. and Qi, Rui and Walker, Brandon D. and Piquemal, Jean-Philip and Ren, Pengyu},
  title = {Polarizable Force Fields for Biomolecular Simulations: Recent Advances and Applications},
  journal = {Annu. Rev. Biophys.},
  year = {2019},
  volume = {48},
  pages = {371--394},
  doi = {10.1146/annurev-biophys-070317-033349}
}

@article{Ventura_2022,
  author = {Ventura, Elizete and Andrade do Monte, Silmar and do Casal, Mariana T. and Pinheiro, Jr., Max and Toldo, Josene Maria and Barbatti, Mario},
  title = {Modeling the heating and cooling of a chromophore after photoexcitation},
  journal = {Phys. Chem. Chem. Phys.},
  year = {2022},
  volume = {24},
  pages = {9403--9410},
  doi = {10.1039/D2CP00686C}
}

@article{Birgisson_2025,
  author = {Birgisson, Benedikt O. and Dohn, Asmus Ougaard and J{\'o}nsson, Hannes and Levi, Gianluca},
  title = {Decoherence and vibrational energy relaxation of the electronically excited {PtPOP} complex in solution},
  journal = {J. Chem. Phys.},
  year = {2025},
  volume = {162},
  number = {4},
  pages = {044306},
  doi = {10.1063/5.0241573}
}

@article{Biasin_2016,
   author = {Elisa Biasin and Tim Brandt Van Driel and Kasper S. Kjær and Asmus O. Dohn and Morten Christensen and Tobias Harlang and Pavel Chabera and Yizhu Liu and Jens Uhlig and Mátyás Pápai and Zoltán Németh and Robert Hartsock and Winnie Liang and Jianxin Zhang and Roberto Alonso-Mori and Matthieu Chollet and James M. Glownia and Silke Nelson and Dimosthenis Sokaras and Tadesse A. Assefa and Alexander Britz and Andreas Galler and Wojciech Gawelda and Christian Bressler and Kelly J. Gaffney and Henrik T. Lemke and Klaus B. Møller and Martin M. Nielsen and Villy Sundström and György Vankó and Kenneth Wärnmark and Sophie E. Canton and Kristoffer Haldrup},
   issn = {10797114},
   issue = {1},
   journal = {Phys. Rev. Lett.},
   month = {6},
   publisher = {American Physical Society},
   title = {Femtosecond X-Ray Scattering Study of Ultrafast Photoinduced Structural Dynamics in Solvated [Co (terpy)2] 2+},
   volume = {117},
   year = {2016}
}

@article{Panman_2020,
   author = {Matthijs R. Panman and Elisa Biasin and Oskar Berntsson and Markus Hermann and Stephan Niebling and Ashley J. Hughes and Joachim Kübel and Kalina Atkovska and Emil Gustavsson and Amke Nimmrich and Asmus O. Dohn and Mads Laursen and Diana B. Zederkof and Alireza Honarfar and Kensuke Tono and Tetsuo Katayama and Shigeki Owada and Tim B. Van Driel and Kasper Kjaer and Martin M. Nielsen and Jan Davidsson and Jens Uhlig and Kristoffer Haldrup and Jochen S. Hub and Sebastian Westenhoff},
   doi = {10.1103/PhysRevLett.125.226001},
   issn = {10797114},
   issue = {22},
   journal = {Phys. Rev. Lett.},
   month = {11},
   pmid = {33315438},
   publisher = {American Physical Society},
   title = {Observing the Structural Evolution in the Photodissociation of Diiodomethane with Femtosecond Solution X-Ray Scattering},
   volume = {125},
   year = {2020}
}

@article{Hansen_2025,
   author = {Bianca L. Hansen and Verena Markmann and Mátyás Pápai and Philipp Lenzen and Morten Lunn Haubro and Antal Mikeházi and Zoltán Németh and Andor Vancza and Matteo Levantino and Serhane Zerdane and Diana Bregenholt Zederkof and Dmitry Khakhulin and Asmus Ougaard Dohn and Martin Meedom Nielsen and Kristoffer Haldrup and György Vankó},
   issn = {10897690},
   issue = {12},
   journal = {J. Chem. Phys.},
   month = {3},
   pmid = {40162540},
   publisher = {American Institute of Physics},
   title = {Excited-state structural characterization of a series of nanosecond-lived [Fe(terpy)2]2+ derivatives using x-ray solution scattering},
   volume = {162},
   year = {2025}
}

@article{Mara_2025,
   author = {Michael W. Mara and Denis Leshchev and Nicholas P. Weingartz and Arnold M. Chan and Adam K. Nijhawan and Andrew J.S. Valentine and Xiaolin Liu and Sarah Kromer and Darren Hsu and Pyosang Kim and Elisa Biasin and Timothy Brandt van Driel and Felix N. Castellano and Xiaosong Li and Lin X. Chen},
   doi = {10.1021/acs.jpclett.5c01245},
   issn = {19487185},
   issue = {28},
   journal = {J. Phys. Chem. Lett.},
   month = {7},
   pages = {7120-7126},
   pmid = {40609039},
   publisher = {American Chemical Society},
   title = {Characterizing Ultrafast Intersystem Crossing Pathways in Molecular Pt Dimers Using Time-Resolved Wide-Angle X-ray Scattering},
   volume = {16},
   year = {2025}
}

@article{Lenzen_2025,
   author = {Philipp Lenzen and Kristoffer Haldrup and Asmus O. Dohn and Frederik Beyer and Elisa Biasin and Morten Christensen and Bianca L. Hansen and Tobias Harlang and Kasper Skov Kjær and Mads Goldschmidt Laursen and Peter Vester and Tim B. van Driel and Matthieu Chollet and James M. Glownia and Robert J. Hartsock and Henrik T. Lemke and Silke Nelson and Sanghoon Song and Kasper Steen Pedersen and Kelly J. Gaffney and Klaus B. Møller and Martin M. Nielsen},
   issn = {23297778},
   issue = {4},
   journal = {Struct. Dyn.},
   month = {5},
   publisher = {American Crystallographic Association},
   title = {Structure and ultrafast dynamics of tri-nuclear Ag-/Tl-Pt2POP4 complexes in solution},
   volume = {12},
   year = {2025}
}

\end{document}


\title{Supporting Information for ``Simulated Solvation Dynamics of a Photoexcited Organic Molecule and the Associated Time-Resolved X-ray Scattering Signal''}
\author{S. Petry}
\email{simon.petry@units.it}
\affiliation{Department of Chemical and Pharmaceutical Sciences, University of Trieste, 34127 Trieste, Italy}
\author{M. Kristinsson}%
\affiliation{Department of Chemistry, Technical University of Denmark, Kongens Lyngby, Denmark.}
\author{A. O. Dohn}
\affiliation{Department of Physics, Technical University of Denmark, Kongens Lyngby, Denmark.}
\author{K. B. M{\o}ller}
\affiliation{Department of Chemistry, Technical University of Denmark, Kongens Lyngby, Denmark.}
\author{E. Selenius}
\affiliation{Science Institute and Faculty of Physical Sciences, University of Iceland, 107 Reykjavík, Iceland}
\author{G. Levi}
\email{gianluca.levi@units.it}
\affiliation{Department of Chemical and Pharmaceutical Sciences, University of Trieste, 34127 Trieste, Italy}
\affiliation{Science Institute and Faculty of Physical Sciences, University of Iceland, 107 Reykjavík, Iceland}

\date{\today}

\setlength{\parindent}{0pt}

\maketitle

\tableofcontents

\begingroup
\makeatletter
\let\toc@pre\relax
\let\toc@post\relax
\makeatother
\listoffigures
\endgroup
\clearpage

\section{Equilibrium radial distribution functions}
\begin{figure*}[!htb]
    \centering    
    \includegraphics[width=\textwidth,height=0.28\textheight,keepaspectratio]{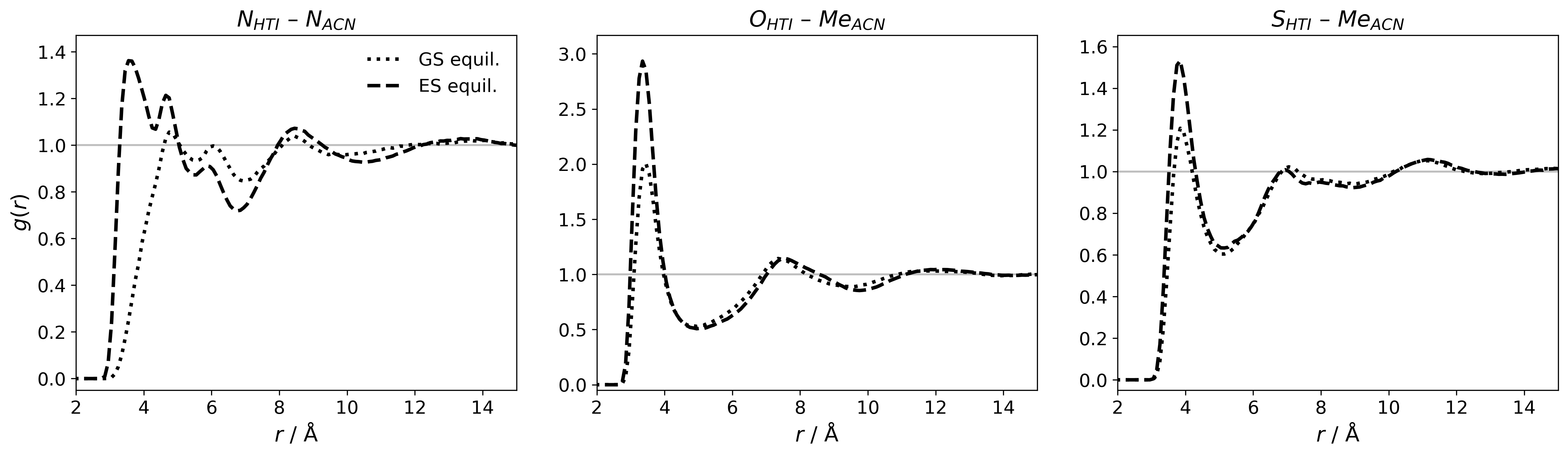}
    \caption[
    Equilibrium GS and ES radial distribution functions for N–N, O–Me and S–Me pairs of HTI in acetonitrile.
    ]{
    Equilibrium radial distribution functions, $g(r)$, of the distance between atoms of the HTI and acetonitrile (ACN) solvent molecules, computed from equilibrated ground-state (GS, dotted) and excited-state (ES, dashed) trajectories. Left: $N_{HTI}–N_{ACN}$. Center: $O_{HTI}–Me_{ACN}$. Right: $S_{HTI}–Me_{ACN}$, where Me denotes the methyl carbon of ACN. The grey horizontal line marks the bulk limit, $g(r) = 1$.
    }
    \label{fig:RDF}
\end{figure*}
\clearpage

\section{Additional Spatial distribution functions}








\begin{figure*}[h]
    \centering    
    \includegraphics[width=\textwidth,height=0.38\textheight,keepaspectratio]{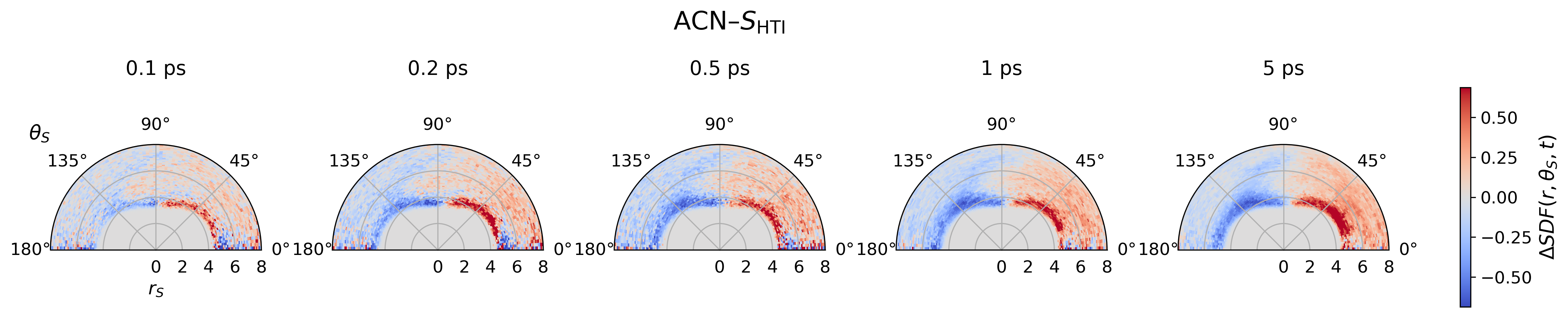}
    \caption[
    Time-resolved difference SDFs around \(S_{\mathrm{HTI}}\)
    ]{\small Time-resolved difference spatial distribution functions (SDFs) of acetonitrile around the sulfur atom \(S_{\mathrm{HTI}}\) at selected delay times after charge-transfer excitation, relative to the equilibrium ground-state distribution.
    The radial coordinate \(r_S\) measures the distance from the sulfur atom to the midpoint between the acetonitrile nitrogen and methyl carbon, while \(\theta_S\) is the angle between the acetonitrile \(N\!\rightarrow C_{\mathrm{Me}}\) axis and the vector from this midpoint toward the sulfur atom.
    }
    \label{fig:SDFs_SI}
\end{figure*}

\begin{figure*}[h]
    \centering
    \includegraphics[width=\textwidth,height=0.38\textheight,keepaspectratio]{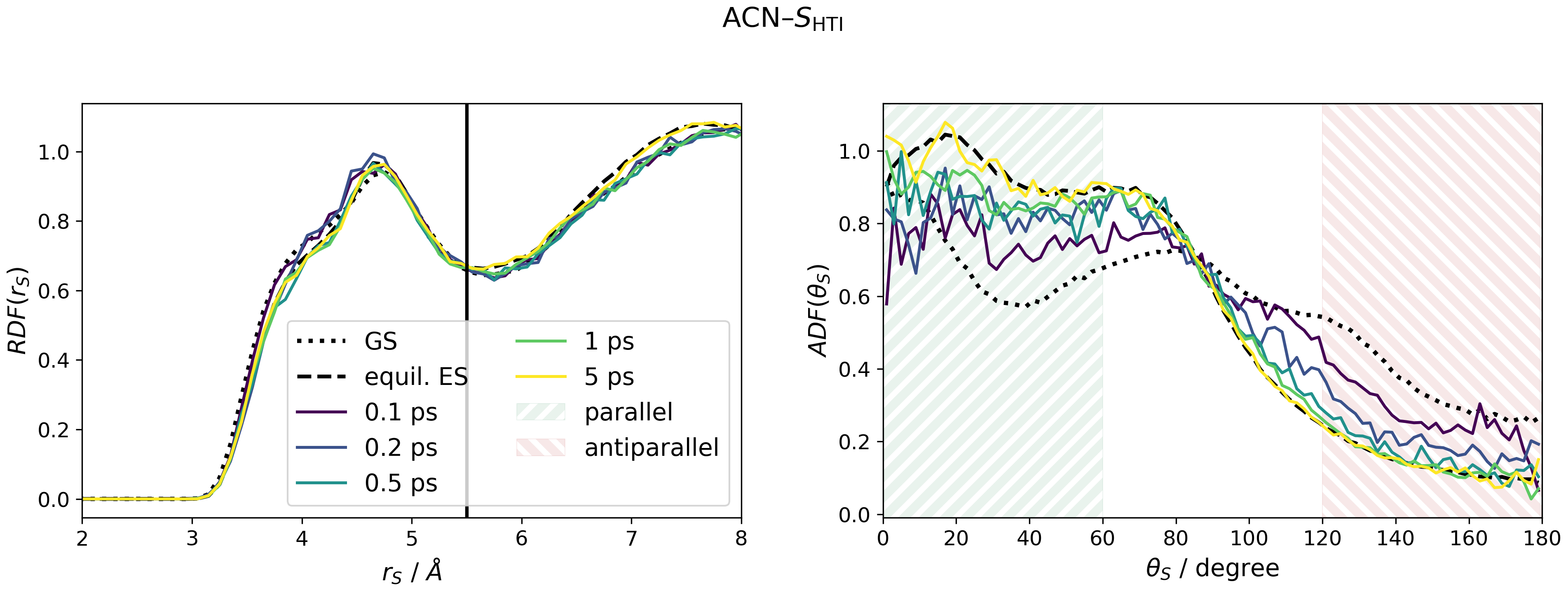}
    \caption[
    Radial and angular SDF projections around \(S_{\mathrm{HTI}}\)
    ]{\small 
    Radial and angular distribution functions of acetonitrile around \(S_{\mathrm{HTI}}\) at selected delay times after photoinduced charge transfer.
    Left: Radial distribution functions, \(g_S(r,t)\), for the distance \(r_S\) between the sulfur atom and the midpoint between the acetonitrile nitrogen and methyl carbon, averaged over all molecular orientations.
    Right: Angular distribution functions, \(a_S(\theta,t)\), obtained by averaging the SDFs up to \(r_S=5.5~\text{\AA}\) (vertical line in the left panel), with \(\theta_S\) being the angle between the acetonitrile \(N\!\rightarrow C_{\mathrm{Me}}\) axis and the vector from its \(N\!-C_{\mathrm{Me}}\) midpoint toward sulfur.
    The hatched areas define parallel (\(0^\circ\leq\theta_S\leq60^\circ\), methyl end toward sulfur) and antiparallel (\(120^\circ\leq\theta_S\leq180^\circ\), nitrile nitrogen toward sulfur) orientation ranges.
    }
    \label{fig:SDF_projections_SI}
\end{figure*}

\begin{figure*}[h]
    \centering
    \includegraphics[width=\textwidth,height=0.38\textheight,keepaspectratio]{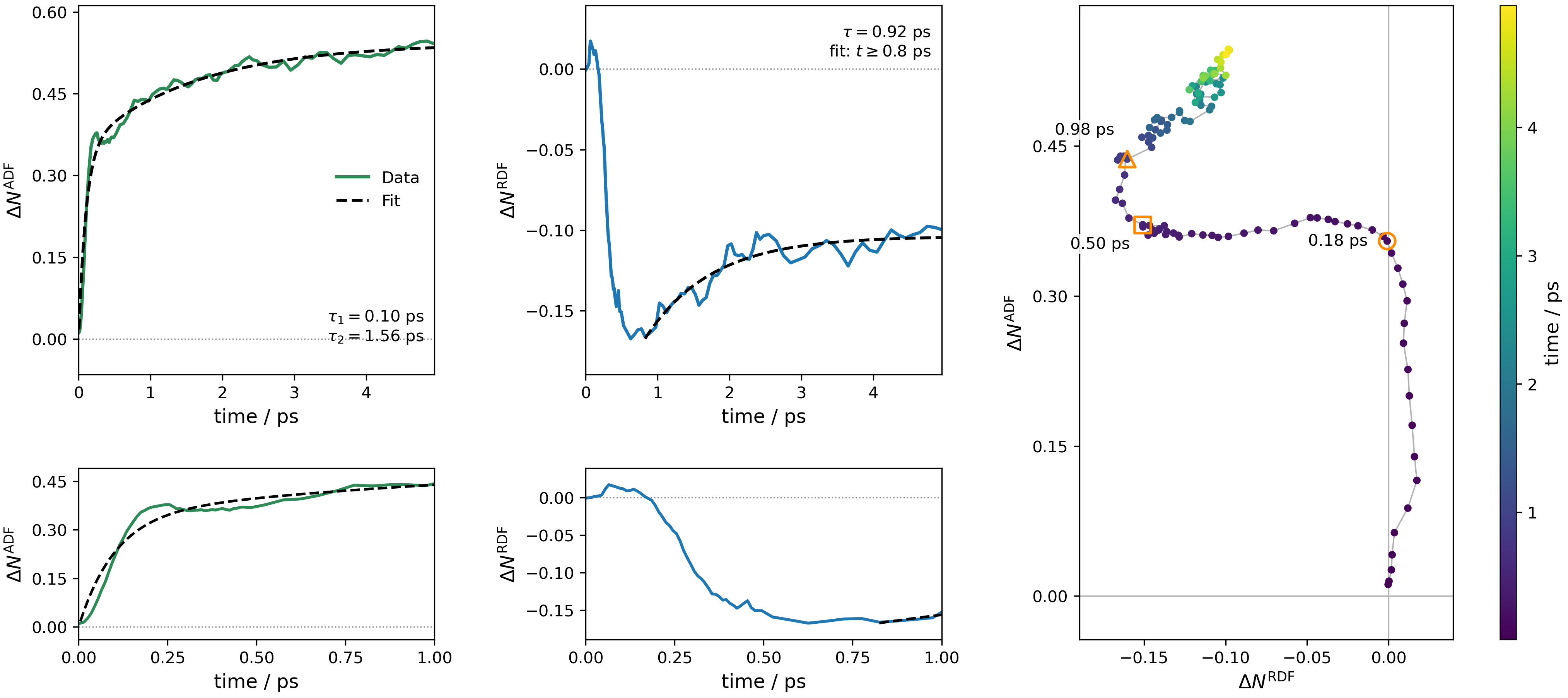}
    \caption[
    Radial and orientational population dynamics around \(S_{\mathrm{HTI}}\)
    ]{
    Time-dependent radial and orientational population changes around \(S_{\mathrm{HTI}}\) following charge-transfer excitation, relative to equilibrium in the ground state.
    Left: change in the number of acetonitrile molecules within \(0\leq r_S\leq5.5~\text{\AA}\) and the parallel range \(0^\circ\leq\theta_S\leq60^\circ\) defined in Fig.~\ref{fig:SDF_projections_SI}.
    Center: Change in cumulative coordination number within the same radial interval.
    The lower panels show enlarged views of the first picosecond.
    Right: Orientational population change plotted against radial population change, with color indicating the delay time after excitation. The orientational population increases rapidly. After a short initial increase, the radial population falls to a minimum at approximately \(0.6\)--\(0.8~\mathrm{ps}\) and subsequently recovers only partially.
    }
    \label{fig:population_dynamics}
\end{figure*}
\clearpage

\section{Additional details of the excluded volume calculation}

Here we provide further details of the excluded volume calculation described in the main text.
For each trajectory frame at delay time \(t\), the surrounding solvent atoms were represented by spheres centered on their instantaneous positions, with effective radii
\begin{equation}
    R_j = r_j^{\mathrm{vdW}} + r_{\mathrm{probe}}.
\end{equation}
Here, \(r_j^{\mathrm{vdW}}\) is the van der Waals radius of atom \(j\), taken from Bondi,\cite{Bondi_1964} and \(r_{\mathrm{probe}}\) is a uniform increment of the radius used to expand the solvent spheres and define the cavity boundary.
For acetonitrile, the van der Waals radii are \(r_{\mathrm{H}}^{\mathrm{vdW}}=1.20~\text{\AA}\), \(r_{\mathrm{C}}^{\mathrm{vdW}}=1.70~\text{\AA}\), and \(r_{\mathrm{N}}^{\mathrm{vdW}}=1.55~\text{\AA}\).

To restrict the calculation to solvent molecules near HTI, the three closest acetonitrile molecules to each solute atom were first selected.
The distance from a solute atom to an acetonitrile molecule was defined as the smallest distance to any atom of that molecule.
For each initially selected molecule, the four nearest additional acetonitrile molecules were also included.
All atoms belonging to the resulting set of solvent molecules were represented by the expanded spheres described above.

The cavity volume was evaluated on a Cartesian grid with spacing \(\Delta x=0.25~\text{\AA}\).
Using the positions of the HTI heavy atoms as starting points, we identified the connected grid cells outside all solvent spheres.
A frame was discarded if this region reached the grid boundary, indicating that the cavity was open to the surrounding solvent.
For a closed cavity, the volume was calculated by multiplying the number of enclosed grid cells by the volume of one cell:
\begin{equation}
    V_{\mathrm{EV}} = N_{\mathrm{cav}}\,\Delta x^3,
\end{equation}
where \(N_{\mathrm{cav}}\) is the number of grid cells belonging to the cavity.

The probe radius \(r_{\mathrm{probe}}=1.60~\text{\AA}\) was selected by testing different values and balancing closure of the cavity with excessive overlap of the expanded solvent spheres with the region occupied by HTI.
With smaller probe radii, gaps between the solvent spheres more frequently connect the cavity to the surrounding solvent, reducing the proportion of closed cavities.
Larger probe radii improve closure but cause the solvent spheres to extend further into the solute region.

\begin{figure*}[h]
    \centering    
    \includegraphics[width=\textwidth,height=0.28\textheight,keepaspectratio]{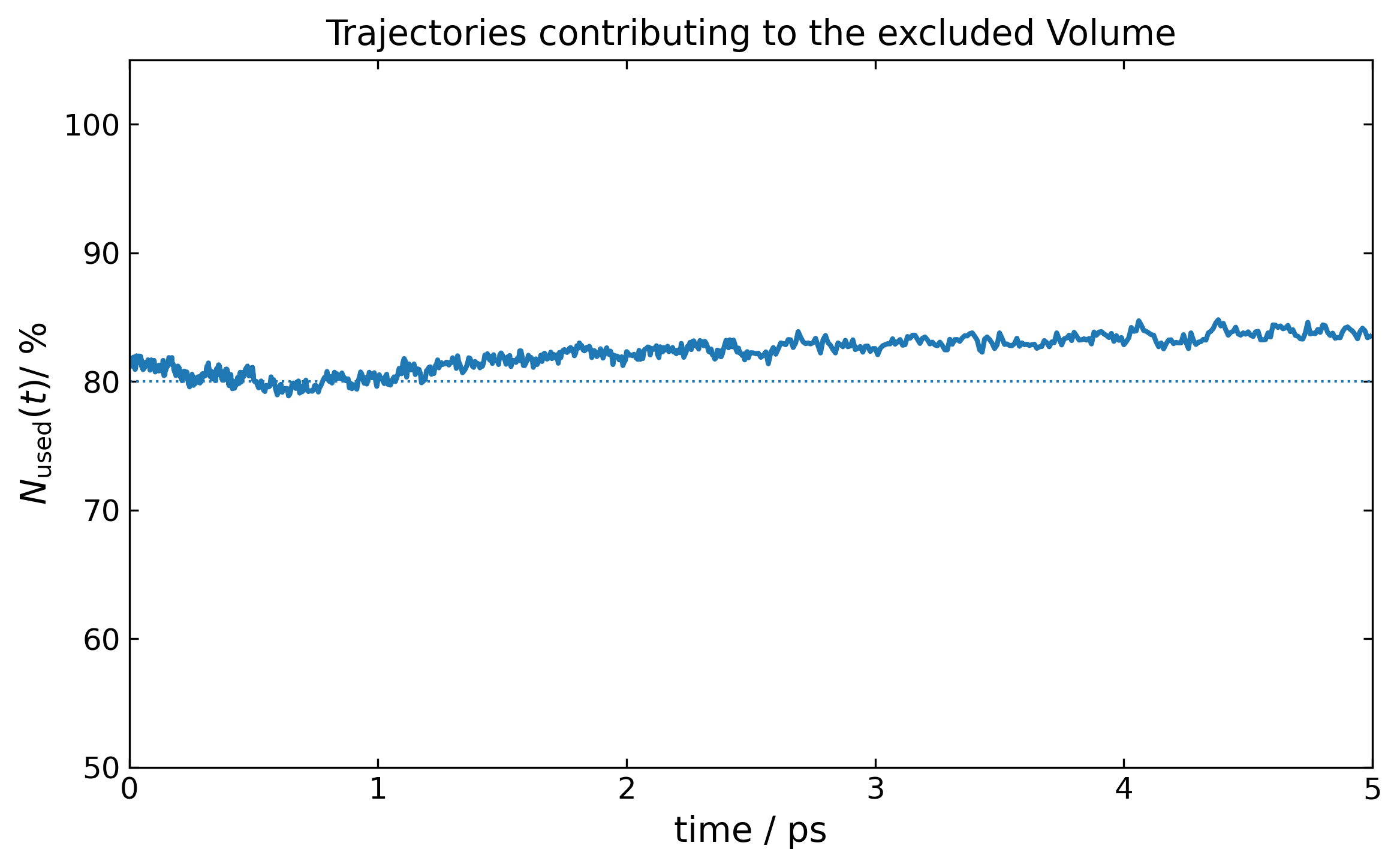}
    \caption[
    Trajectories contributing to the excluded-volume analysis
    ]{
    Proportion of the 5,000 nonequilibrium trajectories of the photoinduced charge-transfer simulations that contribute to the mean value of the excluded volume at each delay time. Only configurations for which a closed cavity associated with HTI could be identified were considered. Over the simulation time, between 78.9\% and 84.8\% of the trajectories contribute to the mean.
    }
    \label{fig:excluded_volume_valid_fraction}
\end{figure*}

The excluded volume was averaged at each delay time over the trajectories for which a closed cavity could be identified:
\begin{equation}
    \left\langle V_{\mathrm{EV}}(t) \right\rangle
    =
    \frac{1}{N_{\mathrm{used}}(t)}
    \sum_{i\in\mathcal{C}(t)} V_{\mathrm{EV},i}(t),
\end{equation}
where \(\mathcal{C}(t)\) is the set of trajectories with a closed cavity at time \(t\), and \(N_{\mathrm{used}}(t)\) is their number.
Of the 5,000 trajectories of the nonequilibrium simulations of photoinduced charge transfer included in the analysis, between 3,944 and 4,240 trajectories (\(78.9\)--\(84.8\%\)) contributed to the ensemble mean, depending on the delay time.
The corresponding fraction of contributing trajectories is shown in Fig.~\ref{fig:excluded_volume_valid_fraction}.

\clearpage

\section{Scattering-like transform of solvent redistribution}

Fig.~6 of the main text shows the distribution of minimum distances, \(r_{\min}\), from the center of mass of each acetonitrile molecule to the closest HTI atom.
Here, \(n(r_k,t)\) denotes the ensemble-averaged number of acetonitrile molecules with minimum distance falling within a distance bin of width \(\Delta r=0.1~\text{\AA}\) centered at \(r_k\).
The population change, \(\Delta n(r_k,t)=n(r_k,t)-n_{\mathrm{GS}}(r_k)\), is evaluated relative to the equilibrium ground-state distribution.

To relate these population changes to the \(q\)-dependence of isotropic scattering, we construct a scattering-like transform.
For a selected minimum distance interval \(r_1\leq r_{\min}<r_2\):
\begin{equation}
    \Delta P(q,t;r_1,r_2)
    =
    \sum_{r_1\leq r_k<r_2}
    \Delta n(r_k,t)\,j_0(qr_k).
    \label{eq:population_transform}
\end{equation}
Here, \(q\) is the magnitude of the scattering vector.
The $j_0(qr_k)$ kernel is the zeroth-order spherical Bessel function, defined by
\begin{equation}
    j_0(qr)=\frac{\sin(qr)}{qr}.
    \label{eq:spherical_bessel_kernel}
\end{equation}
The kernel \(j_0(qr)\) arises from the orientational average of the plane-wave phase factor for a pair of scatterers separated by a distance \(r\).\cite{Warren_1990}
It therefore also appears in the Debye expression for isotropic scattering\cite{Debye_1915} and in its formulation in terms of atom-pair distribution functions.
Eq.~(9) of Dohn \textit{et al.}\cite{Dohn_2015} shows how isotropic averaging reduces the three-dimensional integral over a pair density to a distance integral containing this kernel.

The transform uses the same distance-dependent weighting as the difference scattering calculation, but assigns equal weight to all molecules.
The calculated solute--solvent signal instead includes all solute--solvent atom-pair distances, weighted by products of atomic form factors.
Consequently, \(\Delta P(q,t)\) provides a qualitative connection between solvent redistribution and scattering, rather than a quantitative estimate of \(\Delta S(q,t)\).
In particular, it does not retain the full information about solvent orientation and the distribution of solvent atoms around individual solute sites.

In the limit \(q\rightarrow0\), the kernel approaches unity.
The transform therefore reduces to the sum of the population changes within the selected minimum distance interval:
\begin{equation}
    \Delta P(q,t;r_1,r_2)
    =
    \sum_{r_1\leq r_k<r_2}\Delta n(r_k,t).
    \label{eq:population_transform_zero_q}
\end{equation}
This limit equals the net population change within that interval.

\begin{figure}[t]
    \centering
    \includegraphics[width=0.8\columnwidth]{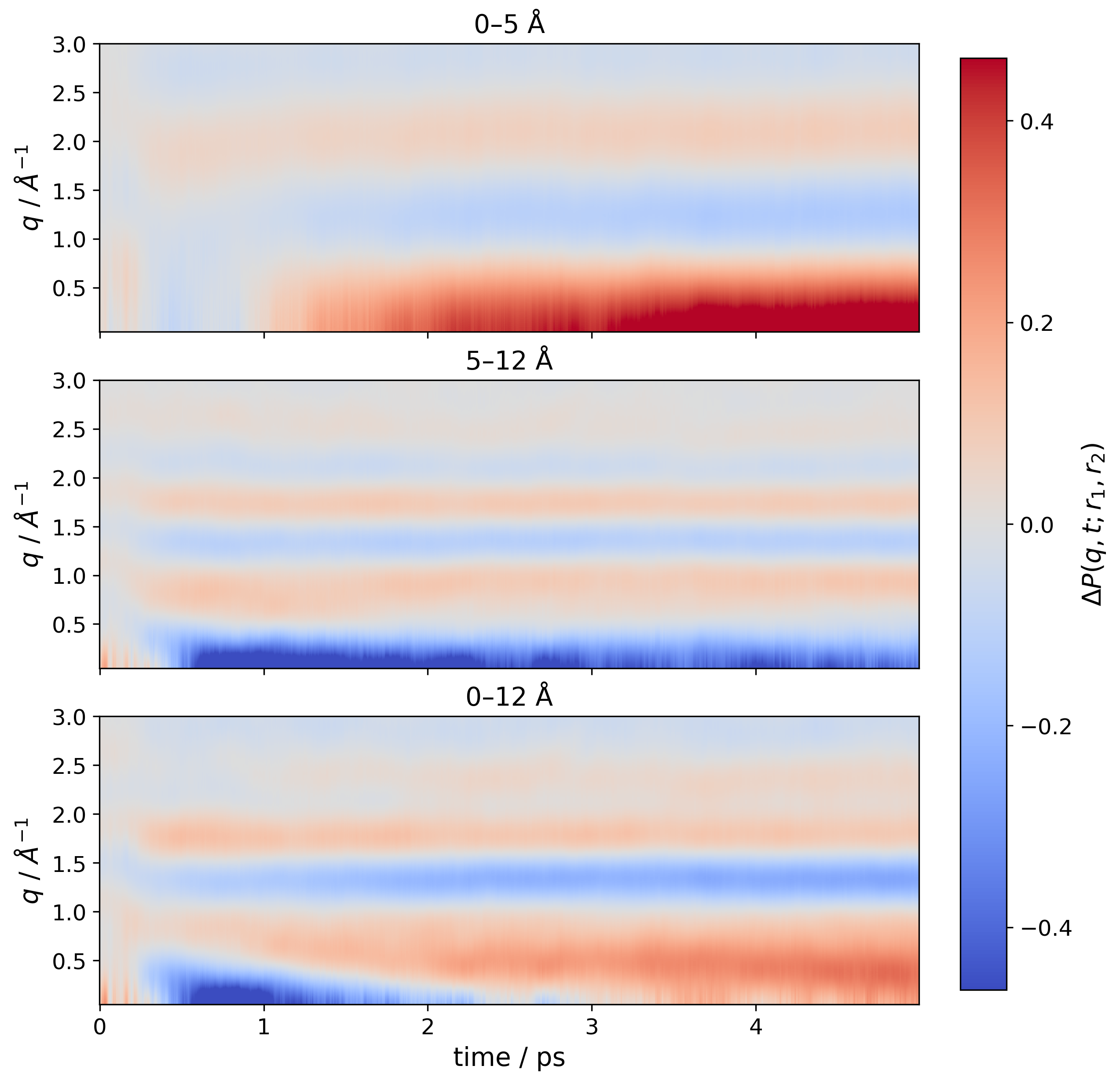}
    \caption[Scattering-like transform of solvent redistribution]{
    Scattering-like transform \(\Delta P(q,t)\), defined in Eq.~\eqref{eq:population_transform}, as a function of the magnitude of the scattering vector, \(q\), and delay time after charge-transfer excitation, relative to the equilibrium ground-state distribution.
    The minimum distance coordinate \(r_{\min}\) measures the distance from the center of mass of each acetonitrile molecule to the closest HTI atom.
    Top: \(0\leq r_{\min}<5~\text{\AA}\).
    Middle: \(5\leq r_{\min}<12~\text{\AA}\).
    Bottom: The full region, \(0\leq r_{\min}<12~\text{\AA}\). The low-$q$ feature closely resembles the low-$q$ part of the solute-solvent scattering signal shown in Fig. 7 of the main text.
    }
    \label{fig:Scattering_map}
\end{figure}

Figure~\ref{fig:Scattering_map} shows the dependence of the transform on \(q\) and delay time for three minimum distance ranges identified in Fig. 6 of the main tex: (1) \(0\leq r_{\min}<5~\text{\AA}\), approximately encompassing the first solvation shell, (2) \(5\leq r_{\min}<12~\text{\AA}\), covering the surrounding solvent layers, and (3) the full \(0\leq r_{\min}<12~\text{\AA}\) region.
The inner and outer contributions are additive by construction, and
can reinforce or partially cancel each other.

The combined transform displays qualitative similarities to the calculated solute--solvent difference scattering signal in Fig.~7 of the main text below approximately \(0.7~\text{\AA}^{-1}\). 
Both show an early negative feature at small \(q\), followed by the development of a positive feature.
This closer correspondence at low \(q\) is consistent with sensitivity to broader changes in solvent packing and redistribution, whereas the detailed atom-pair distances and molecular orientations omitted from the transform become more important at higher \(q\).

The shell-resolved transforms in Fig.~\ref{fig:Scattering_map} indicate that the early negative feature at small \(q\) arises predominantly from the outer region, \(5\leq r_{\min}<12~\text{\AA}\), whereas the growing positive feature originates primarily from the first solvation shell, \(0\leq r_{\min}<5~\text{\AA}\).
This suggests that the negative contribution reflects the initial decrease in the number of molecules in the outer shells followed by partial recovery, while the positive contribution reflects the increase in the first-shell population.

\begin{figure}[t]
    \centering
    \includegraphics[width=\textwidth]{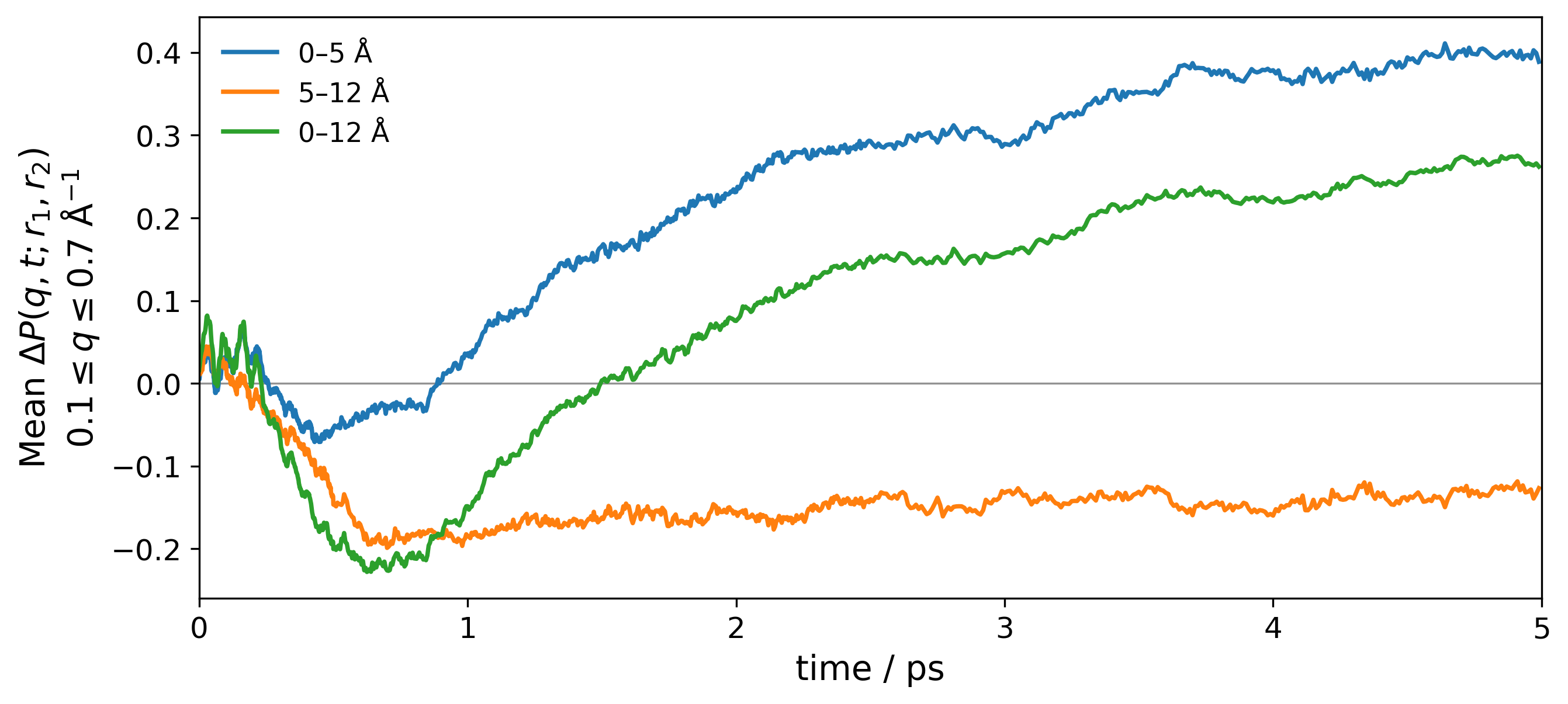}
    \caption[Time evolution of the mean low-\(q\) transform]{
    Time evolution of the scattering-like transform for various minimum distance regions averaged over \(0.1\leq q\leq0.7~\text{\AA}^{-1}\), following charge-transfer excitation and relative to the equilibrium ground-state distribution.
    The minimum distance is measured from the center of mass of each acetonitrile molecule to the closest HTI atom.
    The averaged transform exhibits an initial negative excursion followed by recovery to positive values, closely resembling the time evolution of the low-$q$ averaged solute-solvent difference scattering signal shown in Fig. 8 of the main text.
    }
    \label{fig:Scattering_shell_mean}
\end{figure}

To summarize the temporal evolution across the low-\(q\) interval, Fig.~\ref{fig:Scattering_shell_mean} shows the transform from the various  minimum distance ranges averaged over \(0.1\leq q\leq0.7~\text{\AA}^{-1}\).
The averaged trace emphasizes the initial negative excursion during the first picosecond and the subsequent recovery to positive values.
This evolution resembles the low-\(q\) solute--solvent scattering dynamics shown in Figs.~7 and 8 of the main text, supporting a connection to solvent redistribution within and across the surrounding shells.

\clearpage

\section{Comparison of experimental and simulated scattering from bulk solvent heating}
\begin{figure*}[h]
    \centering    
    \includegraphics[width=\textwidth,height=0.28\textheight,keepaspectratio]{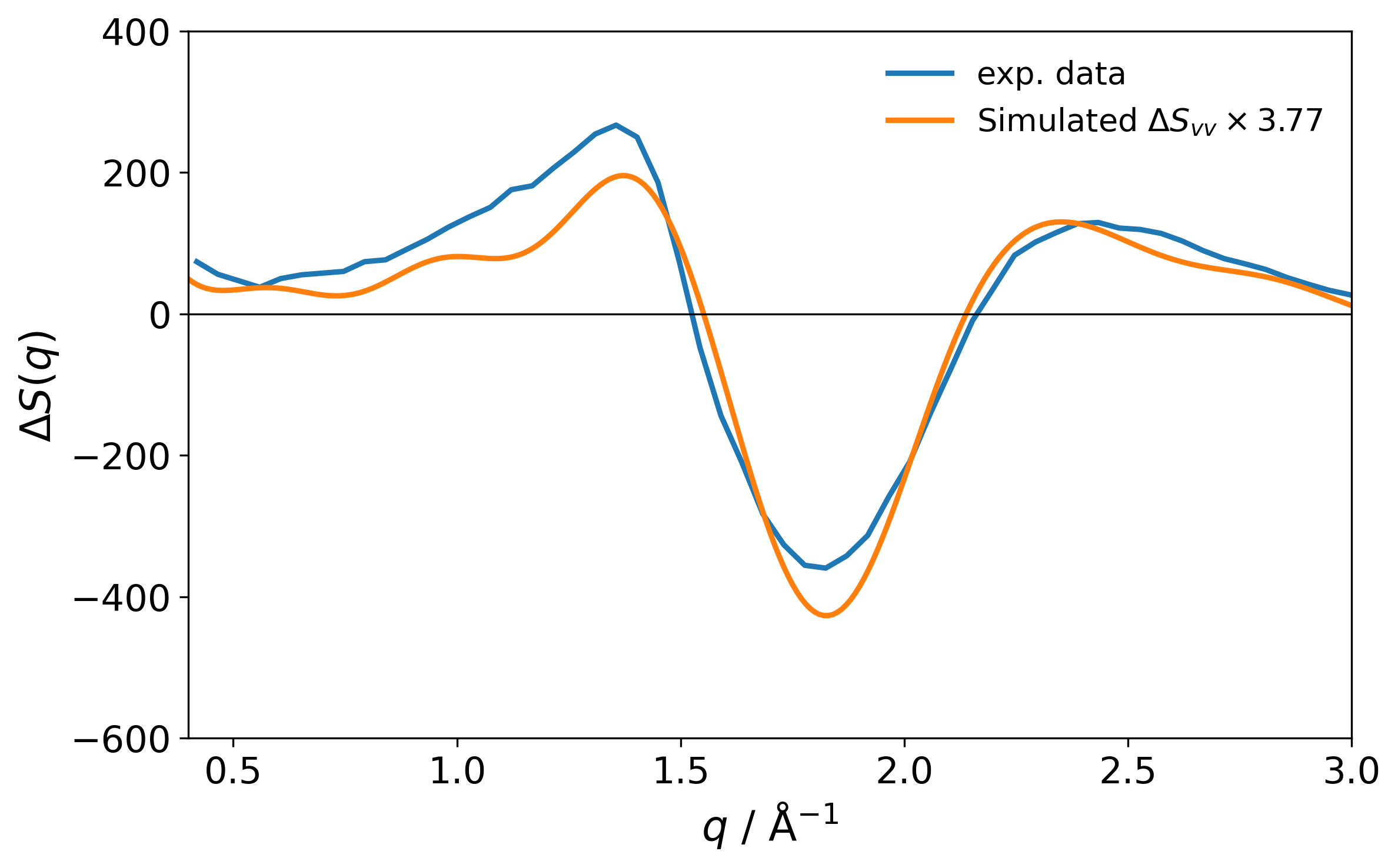}
    \caption[
    Experimental and simulated solvent–solvent scattering signals.
    ]{
    Comparison of the experimental solvent–solvent X-ray scattering signal (blue) with the solvent–solvent contribution obtained from the heating simulations (orange), averaged over 55–60 ps.
    The experimental signal was extracted from the time-resolved X-ray solution scattering data for HTI in acetonitrile reported by Mitterer \textit{et al.}\cite{Mitterer_2026}
    The simulated curve is scaled by 3.77.
    }
    \label{fig:scattering}
\end{figure*}
\clearpage

\section{Kinetic energy evolution after photoinduced charge transfer}
\begin{figure*}[h]
    \centering    
    \includegraphics[width=\textwidth,height=0.38\textheight,keepaspectratio]{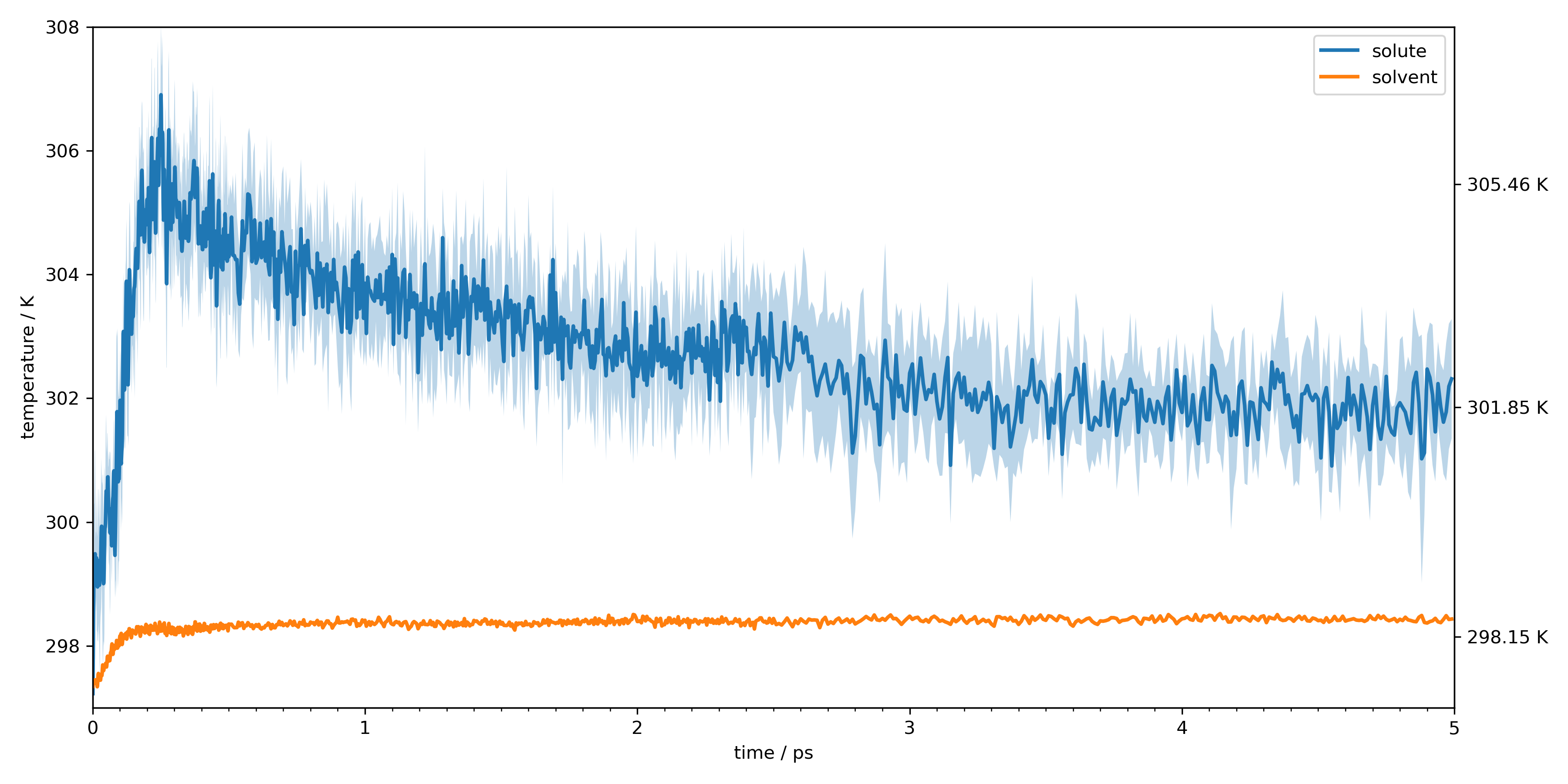}
    \caption[
    Time evolution of solute and solvent kinetic temperatures
    ]{
    Time evolution of the kinetic temperature of the solute (blue) and solvent (orange) following excitation to the charge-transfer state. The light-blue shaded region indicates the statistical spread of the solute temperature.
    }
    \label{fig:energy_SI}
\end{figure*}
\clearpage

\section{Time constants from kinetic fits}
\begin{table}[!ht]
\caption{
Time constants \(\tau_i\) and amplitudes \(A_i\) obtained from exponential fits to the time evolution of solvent population changes, excluded volume, solute and solvent kinetic energies, and solute--solvent difference X-ray scattering traces for the indicated \(q\) intervals.
Each \(A_i\) is the amplitude of an exponential component of the quantity listed in the first column and has the same units as that quantity.
All time constants and the starting point of the fit are in picoseconds.\vspace{0.05cm}
}
\label{tab:timescales}
\centering
\small
\setlength{\tabcolsep}{4pt}
\renewcommand{\arraystretch}{1.05}
\sisetup{group-digits=none}
\begin{tabular*}{\textwidth}{@{\extracolsep{\fill}}ll
    S[table-format=-4.2] S[table-format=1.2]
    S[table-format=-3.2] S[table-format=2.2]
    S[table-format=-1.2] S[table-format=2.2]
    S[table-format=1.2]@{}}
\hline
Quantity & Component / range
& \multicolumn{1}{c}{$A_1$}
& \multicolumn{1}{c}{\shortstack{$\tau_1$}}
& \multicolumn{1}{c}{$A_2$}
& \multicolumn{1}{c}{\shortstack{$\tau_2$}}
& \multicolumn{1}{c}{$A_3$}
& \multicolumn{1}{c}{\shortstack{$\tau_3$}}
& \multicolumn{1}{c}{\shortstack{Fit start}}\\
\hline
\multicolumn{9}{c}{\textbf{Photoinduced charge transfer}}\\
\addlinespace[0.4em]
$\Delta N^{\mathrm{ADF}}$ & $N_{\mathrm{HTI}}$, anti
& 0.83 & 0.18 & 0.60 & 1.15 & {--} & {--} & 0\\
$\Delta N^{\mathrm{RDF}}$ & $N_{\mathrm{HTI}}$
& 0.49 & 1.31 & {--} & {--} & {--} & {--} & 0\\
$\Delta N^{\mathrm{ADF}}$ & $O_{\mathrm{HTI}}$, parallel
& 0.89 & 0.14 & 0.59 & 1.04 & {--} & {--} & 0\\
$\Delta N^{\mathrm{RDF}}$ & $O_{\mathrm{HTI}}$
& 0.27 & 1.40 & {--} & {--} & {--} & {--} & 0\\
$\Delta N^{\mathrm{ADF}}$ & $S_{\mathrm{HTI}}$, parallel
& 0.35 & 0.10 & 0.20 & 1.56 & {--} & {--} & 0\\
$\Delta N^{\mathrm{RDF}}$ & $S_{\mathrm{HTI}}$
& -0.10 & 0.92 & {--} & {--} & {--} & {--} & 0.8\\
\addlinespace[0.3em]
$V_{\mathrm{EV}}$ &
& 10.12 & 1.15 & {--} & {--} & {--} & {--} & 0.39\\
$\Delta N_{\mathrm{ACN}}$ & $0$--$5$~\text{\AA}
& -0.61 & 1.32 & {--} & {--} & {--} & {--} & 0.85\\
$\Delta N_{\mathrm{ACN}}$ & $0$--$12$~\text{\AA}
& -0.90 & 1.66 & {--} & {--} & {--} & {--} & 0.85\\
\addlinespace[0.3em]
Mean $\Delta S(q,t)$ & $0.1 \leq q \leq 0.7~\text{\AA}^{-1}$
& -2497.40 & 1.23 & {--} & {--} & {--} & {--} & 0\\
Mean $\Delta S(q,t)$ & $0.9 \leq q \leq 1.2~\text{\AA}^{-1}$
& -115.87 & 1.53 & {--} & {--} & {--} & {--} & 0\\
\hline
\multicolumn{9}{c}{\textbf{Vibrational excitation}}\\
\addlinespace[0.4em]
$V_{\mathrm{EV}}$ &
& 2.25 & 0.45 & 3.96 & 8.68 & {--} & {--} & 0.39\\
Kinetic energy & Solute
& 0.11 & 0.37 & 0.41 & 5.38 & 0.74 & 14.44 & 0.03\\
Kinetic energy & Solvent
& -0.16 & 0.22 & -0.65 & 5.65 & -0.64 & 20.09 & 0.03\\
\addlinespace[0.3em]
Mean $\Delta S(q,t)$ & $0.1 \leq q \leq 0.7~\text{\AA}^{-1}$
& -194.47 & 1.17 & -522.00 & 13.88 & {--} & {--} & 0.3\\
Mean $\Delta S(q,t)$ & $0.7 \leq q \leq 1.6~\text{\AA}^{-1}$
& 26.53 & 0.86 & 29.40 & 12.20 & {--} & {--} & 0.3\\
\hline
\end{tabular*}
\end{table}

\clearpage

\bibliographystyle{unsrt}
\bibliography{aipsamp}